\documentclass[a4paper,twocolumn,aps,prb,superscriptaddress,showpacs,floatfix]{revtex4-1}
\usepackage{amssymb,amsmath,stmaryrd}
\usepackage{times}

\usepackage{mathtools}
\usepackage{array}
\usepackage{graphicx}
\usepackage{ifpdf}
\usepackage{color} 
\usepackage{url}
\usepackage{hyperref}
\definecolor{darkblue}{rgb}{0.0,0.0,0.3}
\hypersetup{colorlinks,breaklinks,
            linkcolor=blue,urlcolor=blue,
            anchorcolor=blue,citecolor=blue}

\renewcommand{\vec}[1]{\mathbf{#1}}

\mathchardef\mhyphen="0002D

\renewcommand{\Re}{\text{Re}\,}

\renewcommand{\Im}{\text{Im}\,}

\newcommand{\plus}{ {\scriptscriptstyle +}}
\newcommand{\minus}{{\scriptscriptstyle -}}

\newcommand{\ket}{\rangle}
\newcommand{\bra}{\langle}
\newcommand{\bsube}{\begin{subequations}}
\newcommand{\esube}{\end{subequations}}
\newcommand{\be}{\begin{equation}}
\newcommand{\ee}{\end{equation}}
\newcommand{\bea}{\begin{eqnarray}}
\newcommand{\eea}{\end{eqnarray}}

\newcommand{\mB}{\mathcal{B}}
\newcommand{\mC}{\mathcal{C}}

\newcommand{\mK}{\mathcal{K}}
\newcommand{\mL}{\mathcal{L}}
\newcommand{\mP}{\mathcal{P}}
\newcommand{\mS}{\mathcal{S}}
\newcommand{\mT}{\mathcal{T}}
\newcommand{\mU}{\mathcal{U}}
\newcommand{\mTb}{\bar{\mathcal{T}}}
\newcommand{\hmU}{\hat{\mathcal{U}}}

\newcommand{\hH}{\hat{H}}

\newcommand{\hh}{\hat{h}}

\newcommand{\hpsi}{\hat{\psi}}
\newcommand{\hpsid}{\hat{\psi}^{\dagger}}

\newcommand{\br}{\mathbf{r}}

\newcommand{\bx}{\mathbf{x}}
\newcommand{\by}{\mathbf{y}}

\newcommand{\up}{\underline{p}}
\newcommand{\uq}{\underline{q}}

\newcommand{\half}{\frac{1}{2}}

\newcommand{\w}{\omega}

\newcommand{\nn} {\nonumber}

\newcommand{\vphi}{\varphi}

\renewcommand{\S} {\scriptstyle }

\def\a{\alpha}
\def\b{\beta}

\def\t{\tau}
\def\grad{\mbox{\boldmath $\nabla$}}
\begin{document}

\title{Diagrammatic expansion for positive density-response spectra: 
Application to the electron gas}

\author{A.-M. Uimonen}
\affiliation{Department of Physics, Nanoscience Center, University of Jyv{\"a}skyl{\"a}, 
FI-40014 Jyv{\"a}skyl{\"a}, Finland}

\author{G. Stefanucci}
\affiliation{Dipartimento di Fisica, Universit{\`a} di Roma Tor Vergata, 
Via della Ricerca Scientifica 1, 00133 Rome, Italy; and European Theoretical Spectroscopy Facility (ETSF)}
\affiliation{INFN, Laboratori Nazionali di Frascati, Via E. Fermi 40,
00044 Frascati, Italy}

\author{Y. Pavlyukh}
\affiliation{Institut f\"{u}r Physik, Martin-Luther-Universit\"{a}t
  Halle-Wittenberg, 06120 Halle, Germany}

\author{R. van Leeuwen}
\affiliation{Department of Physics, Nanoscience Center, University of Jyv{\"a}skyl{\"a}, 
FI-40014 Jyv{\"a}skyl{\"a}, Finland; and European Theoretical Spectroscopy Facility (ETSF)}
\date{\today}
%*****************************
\begin{abstract}
    
In a recent paper [Phys. Rev. B {\bf 90}, 115134 (2014)] we put forward a diagrammatic
expansion for the self-energy which guarantees the positivity of the spectral
function. In this work we extend the theory to the density response function. We write the generic
diagram for the density-response spectrum as the sum of ``partitions''. In a partition the
original diagram is evaluated using time-ordered Green's functions on the left-half of the
diagram, antitime-ordered Green's functions on the right-half of the diagram and lesser or
greater Green's function gluing the two halves. As there exist more than one way to cut a
diagram in two halves, to every diagram corresponds more than one partition. We recognize
that the most convenient diagrammatic objects for constructing a theory of positive spectra are the
half-diagrams.  Diagrammatic approximations obtained by summing the squares of
half-diagrams do indeed correspond to a combination of partitions which, by construction,
yield a positive spectrum.  We develop the theory using bare Green's functions and
subsequently extend it to dressed Green's functions.  We further prove a connection
between the positivity of the spectral function and the analytic properties of the
polarizability.  The general theory is illustrated with several examples and then applied
to solve the long-standing problem of including vertex corrections without altering the
positivity of the spectrum. In fact already the first-order vertex 
diagram, relevant to the study of gradient expansion, Friedel 
oscillations, etc., leads to spectra which are negative in certain 
frequency domain. We find that the simplest approximation to cure this deficiency  is given
by the sum of the zero-th order bubble diagram, the first-order vertex diagram and a
partition of the second-order ladder diagram. We evaluate this 
approximation in the 3D homogeneous electron gas and show the
positivity of the spectrum for all frequencies and densities.

%*****************************
\end{abstract}
%71.10.-w 	Theories and models of many-electron systems
%31.15.A- 	Ab initio calculations
%73.22.Dj 	Single particle states
%\keywords{}
\maketitle
%*****************************************************************************************
\section{Introduction}
%*****************************************************************************************

Many-body perturbation theory (MBPT) has played an important role in the understanding of
the excitation properties of many-electron systems ranging from molecules to solids.  An
important class of excitations are the neutral excitations in which (in an approximate
physical picture) electrons are excited from occupied to unoccupied 
states. These
excitations can, for instance, be induced by external light fields and indeed the optical
properties of materials, e.g., the index of refraction, are completely determined by
neutral excitations.  For the understanding of the excitation spectrum many-body effects
are of crucial importance as interactions lead to qualitatively new excited states of the
system like plasmons and excitons in solids or the auto-ionizing states in 
molecules.  In many-body theory the neutral excitation
spectrum is obtained from the density response function $\chi$
which can be calculated by diagrammatic methods.  In practice one does not approximate
$\chi$ directly but instead its irreducible part $\mP$, called the
polarizability. The density  response function 
and the polarizability  are related through the integral equation 
$\chi=\mP+\mP v\chi$ where $v$ represents the two-body interaction between the electrons.  The simplest
approximation to $\mP$ is the Random Phase Approximation (RPA) introduced
by Bohm and Pines~\cite{pines_collective_1952} to study plasmons in the electron
gas. However, many physical phenomena, such as excitons, are not 
described within the RPA. More complicated approximations involve typically a
summation of an infinite class of diagrams, which is usually carried 
out with the
Bethe-Salpeter equation~\cite{strinati_application_1988}. For instance, by summing
ladder diagrams with a static screened interaction the excitonic properties of many solids
are well described~\cite{strinati_application_1988}. Nevertheless, 
there are 
several other circumstances for which even  the static ladder 
approximation is not enough. In fact, this approximation only allows for single excitations and therefore double and higher
excitations are not incorporated~\cite{sangalli_double_2011,sakkinen_kadanoffbaym_2012}.
High excitations can significantly contribute to the spectrum of
molecular systems.  This is, for instance, the case in conjugated 
polymers, e.g., the
polyenes, where the lowest lying singlet states have a double-excitation
character~\cite{cave_dressed_2004}.  Also for metallic systems there are several physical
situations that require a theoretical treatment beyond the
RPA\cite{huotari_electron-density_2008, sternemann_correlation-induced_2005}.  
The calculation of plasmon
lifetimes~\cite{huotari_dynamical_2011,cazzaniga_dynamical_2011} is 
one example. Another example is the calculation of the correlation
induced double-plasmon excitations~\cite{huotari_electron-density_2008,
  sternemann_correlation-induced_2005} which has recently been studied both
experimentally and theoretically for the simple metals. We also 
mention that beyond-RPA approximations  have been investigated actively in time-dependent density
functional theory~\cite{ullrich_TDDFT} (TDDFT), a widely used framework 
to study optical properties of molecules and solids.  A key quantity 
within the TDDFT formalism is the so-called exchange-correlation kernel which depends entirely on
corrections beyond the RPA. Several parametrizations of this kernel based on the
homogeneous electron gas exist~\cite{gross_kohn,nifosi_1998,qian_vignale_2002} but their
application to finite systems is problematic and the development of better approximations
is an active research area.  

From the viewpoint of diagrammatic MBPT the theoretical
description of many-body interactions beyond RPA involves the inclusion of vertex
corrections in the Feynman diagrams for the polarizability. Vertex
corrections have been studied in several
works~\cite{glick_high-frequency_1971,kugler_theory_1975,sturm_dynamical_2000,holas_dynamic_1979,devreese_dielectric_1980,brosens_dielectric_1980,awa_dynamical_1982,awa_dynamical_1982-1,brosens_dynamical_1984,engel_wave-vector_1990}
on the homogeneous electron gas. For this system it was found that 
first
order vertex corrections give rise to negative spectral
functions~\cite{holas_dynamic_1979,brosens_dynamical_1984}.  Also for finite systems it
was found that by either restriction to a certain class of
diagrams~\cite{sangalli_double_2011,hellgren_exact-exchange_2009} or 
by truncation to
a certain order in the iteration of the Hedin equations~\cite{schindlmayr_systematic_1998}
the spectral function  can become negative.  This has two important consequences. First of all, it
destroys the physical picture of the photo-absorption spectrum as a probability
distribution and secondly it can lead to a density response function with poles off the
real axis in the complex frequency plane thereby ruining its analytic and causal
properties. Wrong analytic
properties were, for instance, found in a study of atomic
photo-absorption spectra with vertex corrections  
included to first order~\cite{hellgren_exact-exchange_2009}. 
Concomitantly  it was also shown that the absorption spectrum
becomes negative around the energy of the inner-shell transitions.
In this work we prove that 
a density response function with a positive spectrum has the correct
analytic properties. 

The problem of negative spectra lies in the structure of the vertex correction 
and therefore the solution must be sought in the way diagrammatic 
theory is used.  In a
recent paper~\cite{stefanucci_diagrammatic_2014} we showed that a similar problem occurs
in the spectrum of the Green's function (which describes the photo-emission
spectrum rather than the photo-absorption spectrum). For that case we 
solved the problem by introducing the concept of ``half-diagrams'' 
which we then used to construct  
a diagrammatic expansion distinct from MBPT.
In the present work we  show that a similar procedure
works for the polarizability too. 
We apply the theory to derive the lowest order vertex correction 
which preserves the positivity of the spectrum. We also evaluate this 
correction for the
homogeneous electron gas by using a combination of analytical frequency integrations and
numerical Monte-Carlo momentum integrations to evaluate the diagrams.

The paper is divided as follows. In Section \ref{theory} we introduce the spectral function for
the density response function and the polarizability and give a 
diagrammatic proof of their positivity. The basic idea of the proof 
consists in cutting every diagram into two halves in all possible ways 
and in recognizing the half-diagrams as the fundamental object of the 
diagrammatic expansion. Here we also derive the connection between the analytic properties and the
positivity of the spectral function.  In Section \ref{psd-sec} 
we put forward a diagrammatic method to generate approximate 
polarizability with positive spectra. The method is first developed 
using bare Green's functions and then extended to dressed Green's 
functions. In Section \ref{example} we provide some
illustrative examples of positive approximations and show how to turn 
a MBPT approximation into a positive one by adding a minimal set of 
diagrams. We also address the positivity of approximations 
generated through the Bethe-Salpeter equation in 
Section~\ref{bsesec}. In Section~\ref{numsec}
we evaluate the lowest-order vertex correction which yields a positive 
spectrum in the homogeneous electron gas. Our conclusions
and outlooks are drawn in Section~\ref{conclusions}.

%*****************************************************************************************
\section{Theoretical Framework}
%*****************************************************************************************
\subsection{The density response function}
\label{theory}

We study the properties of the reducible and irreducible response function within the
Keldysh Green's function theory.  Although this theory is usually applied in the context
of non-equilibrium physics we have found that it also provides a natural and powerful
framework for the calculation of {\em equilibrium} spectra.

We consider a system of interacting fermions with Hamiltonian
%---------------------------%
\bea
\hH &=& \int d\bx \, \hpsid(\bx) \hh(\bx) \hpsi(\bx)\nn\\
&&+ \half\int d\bx d\bx' \, \hpsid(\bx)\hpsid(\bx')v(\bx,\bx') \hpsi(\bx')\hpsi(\bx),
\label{hamiltonian}
\eea
%---------------------------%
where the field operators $\hpsi$, $\hpsid$ annihilate and create a fermion at
position-spin $\bx = (\br \sigma)$, and $v(\bx,\bx')$ is the Coulomb interaction.  The
one-body part of the Hamiltonian is $\hh(\bx)=
-\frac{\grad^{2}}{2m}+q V(\bx)$, with $V$ the 
scalar potential and $q$ the fermion charge.  Within the Keldysh formalism
the correlators are defined on the time-loop contour $\mC$ going from $-\infty$ to
$+\infty$ ({\em minus}-branch $\mC_-$) and back to $-\infty$ ({\em plus}-branch $\mC_+$).
Operators on the minus-branch are ordered chronologically
while operators on the plus-branch are ordered anti-chronologically.
The Green's function $G(\bx_1z_1,\bx_2 z_2)$ (like any other two-time correlator) with
times $z_{1}$ and $z_{2}$ on the contour embodies four different functions $G^{\a\b}$,
$\a,\b=+/-$, depending on the branch $\mC_-,\mC_+$ to which $z_{1}$ and $z_{2}$
belong.~\cite{stefanucci_nonequilibrium_2013,stefanucci_diagrammatic_2014} For both times on the minus
branch we have the \emph{time-ordered} Green's function $G^{\minus\minus}$ whereas for
both times on the plus branch we have the {\em anti-time-ordered} Green's function
$G^{\plus\plus}$.
The time-ordered and anti-time-ordered Green's functions can be expressed in terms of
$G^{\minus\plus}\equiv G^{<}$ and $G^{\plus\minus}\equiv G^{>}$ as follows (omitting the
dependence on the position-spin variables)
%---------------------------%
\be
G^{\pm\pm}(t_{1},t_{2})=\theta(t_{1}-t_{2})G^{\gtrless}(t_{1},t_{2})
+\theta(t_{2}-t_{1})G^{\lessgtr}(t_{1},t_{2}).
\ee
The four functions $G^{\a\b}$ form the building blocks of the following diagrammatic analysis.

The central object of this work is the density response function $\chi$ defined as the
contour-ordered product of density deviation operators
%---------------------------%
\bea
\chi (\bx_1 z_1,\bx_2 z_2)=-i\bra \mathcal{T}_{\mC} [\Delta\hat{n}_{H}(\bx_1 z_1) 
\Delta\hat{n} _{H}(\bx_2 z_2)]\ket ,\nonumber
\eea
%---------------------------%
where $\Delta\hat{n} (\bx) =\hat{n} (\bx) -\bra \hat{n}(\bx) \ket $, the subscript ``$H$''
implies operators in the Heisenberg picture and $\mathcal{T}_{\mC}$ is the
contour-ordering operator. The average $\bra\ldots\ket$ is performed over the many-body
state of the system.  The greater $\chi^{\plus\minus}\equiv \chi^>$ and lesser
$\chi^{\minus \plus}\equiv \chi^<$ response functions read
%---------------------------%
\bea
\chi^{>} (\bx_1 t_1,\bx_2 t_2)&=&-i \bra \Delta\hat{n}_{H}(\bx_1 t_1) \Delta\hat{n}_{H}(\bx_2 t_2)\ket 
\label{chi>}
\\
\chi^{<} (\bx_1 t_1,\bx_2 t_2)&=&-i \bra  \Delta\hat{n}_{H}(\bx_2 t_2) \Delta\hat{n}_{H}(\bx_1 t_1) \ket 
\label{chi<}
\eea
%---------------------------%
and fulfill the symmetry relation 
\be
i\chi^{\lessgtr}(\bx_1 t_1,\bx_2 t_2)= [i\chi^{\lessgtr}(\bx_2t_2,\bx_1t_1)]^*.
\label{symrel}
\ee
The quantity of interest for the excitation spectrum  is the retarded response function
\begin{eqnarray}
\chi^R (\bx_1 t_1, \bx_2 t_2) &=& - i \theta (t_1-t_2) \langle [ \hat{n}_H (\bx_1 t_1) , \hat{n}_H (\bx_2 t_2)] \rangle \nonumber \\
&=& \theta (t_1-t_2) ( \chi^>-\chi^<) ( \bx_1 t_1, \bx_2 t_2) .\nonumber
\end{eqnarray}

In equilibrium the Green's functions $G^{\a\b}$ as well as the response functions
$\chi^{\a\b}$ depend on the time-difference $t-t'$ and can therefore be conveniently
Fourier transformed with respect to this time-difference. In the remainder of the paper we
often suppress the dependence on spatial and spin indices and regard all quantities as
matrices with one-particle labels. It is easy to verify that the Fourier transform of the
retarded response function is given by
%---------------------------%
\be
\chi^R (\omega) = \int \frac{d\omega'}{2 \pi} \frac{\mB (\w')}{\w-\w' + i \eta}
\label{chiR}
\ee
%---------------------------%
with $\eta$ a positive infinitesimal and spectral function 
%---------------------------%
\be
\mB (\w) = i [\chi^>(\w) - \chi^<(\w)].
\ee
%---------------------------%
The matrix function $\mB$ contains information on the energy of the neutral excitations
and can be measured in optical absorption experiments.  From the relation in
Eq.~\eqref{symrel} we see that $i\chi^> (\w)$ and $i\chi^< (\w)$ 
are self-adjoint and, therefore,
$\mB$ is self-adjoint too.  The matrix function $i \chi^> (\w)$ vanishes for $\w<0$ and its
average is proportional to the probability of absorbing light with frequency $\w$ for
$\w>0$.  In fact, $i \chi^> (\w)$ is a positive semi-definite (PSD) matrix as it follows
directly from the Lehmann representation. Similarly it can be shown that $i\chi^{<}(\w)$
is PSD and vanishes for $\w>0$. Thus  $\mB (\omega)$ is PSD for positive frequencies and negative
semi-definite for negative fequencies.
Another property which follows 
directly from the definitions in Eqs.~(\ref{chi>}) and (\ref{chi<}) 
is the relation $i\chi^{<}(\w)=[i\chi^{>}(-\w)]^{\ast}$. For 
Hamiltonians with time-reversal symmetry, such as  in 
Eq.~(\ref{hamiltonian}), this relation can also be written as 
$i\chi^{<}(\w)=[i\chi^{>}(-\w)]^{\dagger}$ which implies 
that $\mB (\omega)=-\mB
(-\omega)$.
% is only nonzero for positive frequencies 
% and will pick up the contribution from the positive excitation
% energies while the lesser component  $\chi^< (\w)$ is only nonzero for negative frequencies  and 
% will pick up the contribution from the negative excitation energies. The two functions
% are related by $\chi^> (\w)=\chi^< (-\w)$. 
It is worth noticing for the present work that the PSD property is not guaranteed in
diagrammatic approximations to the response function.  How to construct PSD diagrammatic
approximations is the topic of the next section.

%*****************************************************************************************
\subsection{Positivity of the exact response function}
%*****************************************************************************************

The PSD property of the exact $i \chi^\lessgtr (\w)$ is manifest from the Lehmann
representation of this quantity. It is instead less obvious to prove the PSD property from
the diagrammatic expansion. Here we provide such a proof and bring to light a diagrammatic
structure which forms the basis of a general scheme to construct PSD approximations (or to
turn non PSD approximations into PSD ones by adding a minimal set of diagrams).  We follow
the same line of reasoning as in the recently published work on the PSD property of the
self-energy.~\cite{stefanucci_diagrammatic_2014} The main difference is that the proof for
the response function involves intermediate states consisting of particle-hole pairs
rather than particle-hole pairs plus a particle or a hole.  Since the derivation is
otherwise similar we only outline the basic steps and refer to
Ref.~\onlinecite{stefanucci_diagrammatic_2014} for more details.  Moreover since the
derivation for $\chi^>$ is essentially the same as for $\chi^<$ we restrict the attention
to $\chi^<$.
%
%
%
%One of the differences to our previous paper with the self-energy~\cite{positivity_sigma} is that 
%the polarisability is given in terms of bosonic propagators 
%{\color{red} WHAT IS MEANT HERE?} and the greater and lesser components will have the same
%expression.  
%In order to prove that the spectra given by $\chi_{\mu\nu}^{<}(z_1,z_2)$ is
%positive semidefinite (PSD) we will consider for simplicity a system at zero
%temperature {\color{red}THIS POINT IS SIMILAR TO THE SELF-ENERGY 
%PAPER. THE POSITIVITY OF P IS IMPLIED BY THE POSITIVITY OF CHI WHICH 
%FOLLOWS TRIVIALLY FROM LEHMANN. HERE WE COULD SAY THAT WE WILL 
%PROVIDE AN ALTERNATIVE DIAGRAMMATIC PROOF OF THE POSITIVITY OF *P* 
%WHICH WILL BE USEFUL FOR CONSTRUCTING PSD APPROXIMATIONS }. 

\begin{figure}[b!]
\centering
       \includegraphics[width=\linewidth]{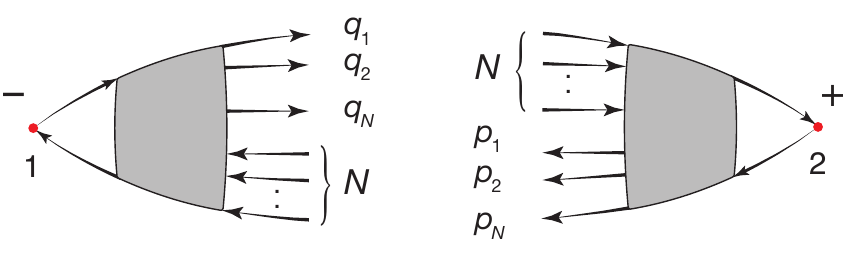}
\caption{(Color online) Typical diagram for $S^*(1)$ (left) and 
$S(2)$ (right) forming the lesser reducible response function. Labels 
$p$ and $q$ on the arrows denote quantum numbers of particles and 
holes respectively, see Eqs.~(\ref{s1s2}). 
\label{fig:ss}}
\end{figure}

The starting point is Eq. (\ref{chi<}) for $\chi^<$. Writing explicitly the time-evolution
operator $\hat{\mU}$ in $\Delta \hat{n}_{H}$ we get
%---------------------------%
\bea
&&i\chi^{<}(1,2) 
\nn \\
&&=\bra \Psi_0 | \hat{\mU}(t_0,t_2) \Delta \hat{n} (\bx_2)\hat{\mU}(t_2,t_1)
\Delta\hat{n} (\bx_1)\hat{\mU}(t_1,t_0)  | \Psi_0 \ket, \nonumber
\eea
%---------------------------%
where $| \Psi_0 \ket$ is the non-degenerate ground state and the short-hand notation
$1=\bx_1 t_1$ and $2=\bx_2 t_2$ has been introduced.
%%---------------------------%
%\bea
%\hat{\mU}(t_1,t_2)=\begin{cases}
%\mT \exp \bigg( -i\int_{t_1}^{t_2}d\tau \hH(\tau) \bigg) & t_2>t_1 \nn\\
%\mTb \exp \bigg( i\int_{t_2}^{t_1}d\tau \hH(\tau) \bigg) & t_2<t_1, \nn\\
%\end{cases}
%\eea
%%---------------------------%
%where $\mT$ is the time-ordering operator arranging earliest time to the left and $\mTb$
%is the anti-time-ordering operator arraying earliest time to the right.  
Under the adiabatic assumption the state $|\Psi_0 \ket = \hat{\mU}(t_0,\tau) | \Phi_0\ket$
is obtained by propagating the noninteracting ground-state $| \Phi_0\ket$ from the distant
future time $\t$ (eventually we take $\tau \to\infty$) to some finite time $t_{0}$ with an
adiabatically switched-on interaction. Therefore
%---------------------------%
\bea
i\chi^{<}(1,2)
&=&\bra \Phi_0 | \hat{\mU}(\tau,t_2) \Delta\hat{n} (\bx_2)\hat{\mU}(t_2,\tau)
\sum_{i}
| \vphi_i \ket\bra \vphi_i |  \nn\\
&&\times \hat{\mU}(\tau,t_1) \Delta\hat{n} (\bx_1) \hat{\mU}(t_1,\tau) | \Phi_0 \ket,
\label{chi_less}
\eea
%---------------------------%
where we inserted the completeness relation in Fock space $\sum_{i} | \vphi_i \ket\bra
\vphi_i |=1$. The only states in Fock space which contribute in Eq.~\eqref{chi_less} have
the form
%---------------------------%
\be
|\vphi_{\up\uq}^{(N)}\ket\equiv 
\hat{c}_{q_N}^{\dagger}....\hat{c}_{q_1}^{\dagger}\hat{c}_{p_N}..\hat{c}_{p_1} | \Phi_0 \ket,
\nn
\ee
%---------------------------%
where $\hat{c}_k$, $\hat{c}_k^{\dagger}$ annihilate and create a fermion in the $k$-th
eigenstate of the noninteracting problem.  In Eq. (\ref{chi_less}) we can therefore
replace
%---------------------------%
\bea
\sum_{i} | \vphi_i \ket \bra \vphi_i |
\rightarrow \sum_{N=1}^{\infty} \frac{1}{N!} \frac{1}{N!}
\sum_{\up\uq} | \vphi_{\up\uq}^{(N)} \ket \bra \vphi_{\up\uq}^{(N)} |,
\label{cr}
\eea
%---------------------------%
where $\sum_{\up\uq}$ denotes integration or summation over the sets
$\up=(p_{1},\ldots,p_{N})$ of occupied states and $\uq=(q_{1},\ldots,q_{N})$ of unoccupied
states.  The sum starts from $N=1$ since the only state with $N=0$ particle-hole pairs is
$|\Phi_0\rangle$ and $\bra\Phi_{0}|\Delta \hat{n}|\Phi_{0}\ket=0$.  The prefactor in
Eq. (\ref{cr}) originates from the inner product of the intermediate states
\begin{equation*}
\bra\vphi_{\up\uq}^{(N)}|\vphi_{\up'\uq'}^{(N')}
\ket=\delta_{N,N'}\sum_{P,Q\in\pi_{N}}(-)^{P+Q}\delta_{P(\up),\up'}\delta_{Q(\uq),\uq'},
\end{equation*}
where $P$ and $Q$ run over the set $\pi_N$ of all possible permutations of $N$ indices 
and  $(-)^P$ and $(-)^Q$ are the parities of the permutations $P$ and 
$Q$ respectively.  Using Eq. (\ref{cr}) in Eq. (\ref{chi_less})
we can rewrite the lesser response function  as
%---------------------------%
\bea
\label{chi_lesser_ss}
i\chi^<(1,2)
&=& \sum_{N=1}^{\infty}  \frac{1}{N!} \frac{1}{N!}\sum_{\up\uq} 
\mS^{(N)}_{\up\uq}(2) \mS^{(N)*}_{\up\uq}(1),
\eea
%---------------------------%
where the amplitudes $\mS$  read
%---------------------------%
\bea
\mS^{(N)*}_{\up\uq}(1) &= \bra \vphi_{\up\uq}^{(N)}| 
\hat{\mU}(\tau,t_1) \Delta\hat{n} (\bx_1) \hat{\mU}(t_1,\tau) | \Phi_0 \ket, \\
\mS^{(N)}_{\up\uq}(2) &= \bra \Phi_0 | \hat{\mU}(\tau,t_2) \Delta\hat{n} (\bx_2) \hat{\mU} (t_2,\tau) 
| \vphi_{\up\uq}^{(N)} \ket.
\eea
%---------------------------%
The adiabatic assumption implies that turning the interaction slowly on and off the state
$\Phi_{0}$ changes at most by a phase factor: $ \hmU(\tau,-\tau) | \Phi_0
\ket=e^{i\alpha}| \Phi_0 \ket$. Hence we can rewrite the amplitudes 
as (for more details see Ref.~\onlinecite{stefanucci_diagrammatic_2014})
%---------------------------%
\begin{widetext}
\begin{subequations}
\label{s}
\bea
\label{s1}
\mS_{\up\uq}^{(N)*}(\bx_1 t_1)
&=& \frac{\bra \Phi_0 | \mT \big\{  e^{-i\int_{-\tau}^{\tau} d\bar{\tau} \hH(\bar{\tau})}
\hat{c}_{p_1}^{\dagger}(\tau^{+})\ldots
\hat{c}_{p_{N}}^{\dagger}(\tau^{+})\hat{c}_{q_1}(\tau)\ldots \hat{c}_{q_N}(\tau)\Delta\hat{n} (\bx_1t_1) \big\} |\Phi_0 \ket}
{\bra \Phi_0 | \mT \big\{  e^{-i\int_{-T}^{T} d\bar{\tau} \hH(\bar{\tau})}  \big \} | \Phi_0 \ket} \\
\label{s2}
\mS_{\up\uq}^{(N)}(\bx_2t_2)&=& \frac{\bra\Phi_0 | \mTb  \big\{  e^{i\int_{-\tau}^{\tau} d\bar{\tau} \hH(\bar{\tau})}
\Delta\hat{n}(\bx_2t_2) \hat{c}_{q_N}^{\dagger}(\tau)\ldots 
\hat{c}_{q_1}^{\dagger}(\tau)\hat{c}_{p_{N}}(\tau^+)\ldots 
\hat{c}_{p_1}(\tau^+) \big\}| \Phi_0 \ket}{
 \bra \Phi_0 | \mTb \big\{  e^{i\int_{-\tau}^{\tau} d\bar{\tau} \hH(\bar{\tau})}  \big \} | \Phi_0 \ket} ,
\eea
\label{s1s2}
\end{subequations}
\end{widetext}
with $\mT$ and $\mTb$ the time-ordering and anti-time-ordering operators respectively. The
time argument in the fermion creation and annihilation operators specifies the position of
the operators on the time axis, and $\tau^+$ denotes a time infinitesimally larger than
$\tau$.  Equations (\ref{s1}) and (\ref{s2}) show that $\mS^{(N)*}$ is an interacting
time-ordered $(N+1)$-Green's function whereas $\mS^{(N)}$ is an interacting
anti-time-ordered $(N+1)$-Green's function.  Hence they can be expanded diagrammatically
using Wick's theorem.\cite{van_leeuwen_wick_2012} The general structure of a $\mS$, $\mS^{\ast}$ diagram is
illustrated in Fig.~\ref{fig:ss} and resembles half a $\chi$ diagram.  The left-half
corresponds to $\mS^{(N)*}$ with lines given by noninteracting time-ordered Green's
functions $g^{--}$ whereas the right-half corresponds to $\mS^{(N)}$ with lines given by
noninteracting anti-time-ordered Green's functions $g^{++}$ (here and in the following we
use the letter $g$ to denote the noninteracting Green's function).

%---------------------------%
%---------------------------%

It is now easy to show that $i\chi^{<}(\w)$ is PSD. By Fourier transforming $\mS$ with
respect to $t_{2}$ and $\mS^*$ with respect to $t_{1}$ we find (omitting the dependence on
the spatial and spin variables)
%---------------------------%
\bea
\label{ft}
i\chi^{<}(t_{1},t_{2}) &=& \sum_{N=1}^{\infty} \frac{1}{N!N!} \int \frac{d\w}{2\pi}\frac{d\w'}{2\pi}\nn\\
&\times&e^{-i\w t_2+i\w' t_1} \sum_{\up\uq}  \mS_{\up\uq}^{(N)} 
(\w)\mS_{\up\uq}^{ (N)*} (\w').
\eea
%---------------------------%
In equilibrium $\chi^{<}$ is invariant under time translations, i.e., it depends on
$t_1-t_2$ only. Imposing time translational invariance on the r.h.s. leads to
%---------------------------%
\be
\sum_{N=1}^{\infty}\frac{1}{N!N!}\sum_{\up\uq}  \mS_{\up\uq}^{(N)} 
(\w)\mS_{\up\uq}^{(N)*} (\w') = 
\mathcal{F} (\w)\delta(\w-\w')
\label{ss=f}
\ee
%---------------------------%
with $\mathcal{F}$ some matrix function of the frequency $\w$. Since for $\w=\w'$ the
l.h.s. is a sum of PSD matrices we conclude that $\mathcal{F}$ is PSD. Inserting
Eq.~(\ref{ss=f}) back into Eq.~\eqref{ft} we see that $\mathcal{F} (\w)$ is the Fourier
transform of $i\chi^{<}$ which, therefore, is PSD too.

\subsection{Positivity of the exact polarizability}
\label{posp}

In MBPT  it is often more convenient to calculate the irreducible 
part $\mP$ of $\chi$ defined by the equation
%---------------------------%
\bea
\label{dyson}
\chi (z_1,z_2) &=& \mP (z_1,z_2) \nonumber \\
 &+& \int_\mC dz dz' \, \mP (z_1,z) v(z,z') \chi (z',z_2) ,
\eea
%---------------------------%
where all quantities are matrices in position-spin space and matrix product is implied.
The two-body interaction in Keldysh space is given by $v(z,z')=v \delta (z,z')$.
Diagrammatically $\mP$ is obtained by removing from $\chi$ all 
diagrams that can be separated into
two pieces by cutting a single interaction line. 
Like the retarded response function in Eq.~(\ref{chiR}) the retarded
polarizability has the spectral representation 
%---------------------------%
\be
\mP^R (\omega) = \int \frac{d\omega'}{2 \pi} \frac{\tilde{\mB} (\w')}{\w-\w' + i \eta}
\label{PR_rep}
\ee
with spectral function
\be
\tilde{\mB} (\w) = i [\mP^>(\w) - \mP^<(\w)].
\ee
The matrix functions $\mP^\lessgtr$ and $\chi^\lessgtr (\w)$ are related by
%---------------------------%
\be
\chi^{\lessgtr} (\w) = [1-\mP^R (\w)v]^{-1} \mP^{\lessgtr} (\w)[1-v 
\mP^A(\w)]^{-1},
\ee
%---------------------------%
where the advanced polarizability $\mP^A(\w)=[\mP^R(\w)]^{\dag}$.
As the exact $\chi^\lessgtr$ is PSD this relation implies that $\mP^\lessgtr$ is PSD
too since $[1-\mP^R v]^\dagger = [1-v \mP^A]$. 

Alternatively we can use the diagrammatic expansion of the exact polarizability to show
that $\mP$ has the PSD property.  In fact, the diagrammatic PSD proof for $\chi$ can
easily be adapted for $\mP$. The removal of (interaction-line) reducible diagrams from
$\chi$ is equivalent to the removal of $\mS$-diagrams that can be separated into a piece
containing the external $\chi$-vertex (either 1 or 2) and a piece containing the $\up \uq$
vertices by cutting a single interaction line. We refer to these $\mS$-diagrams as
irreducible $\mS$-diagrams or irreducible half-diagrams. It is easy to realize that the
product between two irreducible half-diagrams yields an irreducible $\chi$ diagram. If we
define $\tilde{\mS}$ as the sum of irreducible half-diagrams then the lesser polarizability
can be written as
%---------------------------%
\bea
\label{P_lesser_ss}
i\mP^<(1,2)
&=& \sum_{N=1}^{\infty}  \frac{1}{N!} \frac{1}{N!}
\sum_{\up\uq} \tilde{\mS}^{(N)}_{\up\uq}(2) \tilde{\mS}^{(N)*}_{\up\uq} (1).
\eea
%---------------------------%
Fourier transfroming $\tilde{\mS}^{*}$ with respect to $t_{1}$ and $\tilde{\mS}$ with
respect to $t_{2}$, and carrying out the same analysis as in the case of $\chi$, see
Eq. (\ref{ft}), we find that
%---------------------------%
\be
\sum_{N=1}^{\infty}\frac{1}{N!N!}\sum_{\up\uq}  
\tilde{\mS}_{\up\uq}^{(N)} (\w)\tilde{\mS}_{\up\uq}^{ (N)*} (\w') = 
\tilde{\mathcal{F}} (\w)\delta(\w-\w'),
\ee
%---------------------------%
where $\tilde{\mathcal{F}} (\w)$ is a PSD matrix function of the frequency $\w$.  Since
$\tilde{\mathcal{F}}$ is the Fourier transform of $i\mP^<$ then $i\mP^<$ is PSD too.

\subsection{Partitions and cutting rule}

For the subsequent development of a PSD diagrammatic theory we need to introduce the
concept of {\em partitions} of a $\mP^{<}$-diagram. This concept naturally arises when we
multiply two half diagrams, as we shall show below. Due to the anticommuting nature of the
fermionic operators we see from Eq.~\eqref{s} that a permutation $P$ of the $\up$ labels
and a permutation $Q$ of the $\uq$ labels changes the sign of $\mS^{(N)}_{\up\uq}$ (and
hence of $\tilde{\mS}^{(N)}$) by a factor $(-)^{P+Q}$. Let then $\{ D_{\up\uq}^{(j)}\}$ with
$j\in I_N$ be the set of topologically inequivalent diagrams for
$\tilde{\mS}^{(N)}_{\up\uq}$ that are not related by a permutation of 
the $\up$ and $\uq$
labels. By construction we have the expansion
%---------------------------%
\bea
\tilde{\mS}^{(N)}_{\up\uq} = \sum_{j\in I_N}\sum_{P,Q\in \pi_{N}} (-)^{P+Q} D^{(j)}_{P(\up)Q(\uq)}.
\eea
%---------------------------%
Inserting this expansion in Eq.~\eqref{P_lesser_ss} and using the fact that $\pi_{N}$ is a
group we get (for more details see Ref.~\onlinecite{stefanucci_diagrammatic_2014})
%---------------------------%
\bea
i\mP^<(1,2) &=& \sum_{N=1}^{\infty} \sum_{j_1,j_2\in I_N}\sum_{P, Q \in \pi_N}(-)^{P+Q} \nn\\
&\times& \sum_{\up\uq} 
D_{\up\uq}^{(j_2)}(2)D^{(j_1)^*}_{P(\up)Q(\uq)}(1).
\label{pwithpart}
\eea
%---------------------------%
Next we observe that the in- and out-going Green's functions with labels $\up$ and $\uq$
are evaluated at the largest time $\tau$ and, therefore, they are either a greater or a lesser
Green's function. At zero temperature the noninteracting lesser Green's function $g^<$
satisfies the property
%---------------------------%
\be
\sum_p g^<_{\bx p} (t_x,\tau) g^<_{p\by} (\tau,t_y) = i g^<_{\bx \by} (t_x,t_y)
\label{cutting}
\ee
%---------------------------%
with a similar relation for the greater Green's function. Hence, in the sum $
\sum_{\up\uq} D_{\up\uq}^{(j_2)}(2)D^{(j_1)^*}_{P_1(\up)Q_1(\uq)}(1)$ we can replace the
product of two $g^{\lessgtr}$ with a single $g^{\lessgtr}$ connecting two internal
vertices. The result is a polarizability diagram in
which the lines of the left-half are time-ordered Green's functions, the lines of the
right-half are anti-time-ordered Green's functions and the lines connecting the two halves
are either lesser or greater Green's functions.  To represent this type of diagrams we
label every internal vertex with a $+$ or a $-$ and introduce the graphical rule according
to which a line connecting a vertex with label $\a=\pm$ to a vertex with label $\b=\pm$ is
a $g^{\a\b}$. Let us name {\em partition} a $\mP^{<}$ diagram with decorated $\pm$
vertices. The full $\mP^{<}$ diagram is given by the sum of all partitions. The reverse
operation of splitting a partition into two half-diagrams consists in cutting all the
Green's function lines between $-$ and $+$ vertices. In the following we refer to this
reverse operation as the {\em cutting rule}.

Before concluding this section we notice that in Keldsyh formalism a $\mP^<(t_{1},t_{2})$
diagram is obtained from the corresponding Keldysh diagram $\mP(z_{1},z_{2})$ by placing
the contour time $z_{1}$ on the {\em minus}-branch $\mC_-$, the contour time $z_{2}$ on
the {\em plus}-branch $\mC_+$ and by integrating every internal contour time over
$\mC$. Thus the $\mP^{<}$ diagram is the sum of diagrams with internal vertices decorated
in all possible ways.  Our derivation shows that decorated diagrams which fall into
multiple disjoint pieces by cutting the lines between $-$ and $+$ vertices do not
contribute, i.e., they sum up to zero. In fact, these diagrams cannot be written as the
product of two half-diagrams.

%*****************************************************************************************
\subsection{Connection between positivity and analytic properties}
\label{analytic-sec}
%*****************************************************************************************

It follows from the Lehmann representation that the retarded density response function
$\chi^R (\omega)$ has poles at $\pm \Omega_j - i\eta$ where $\Omega_j$ are the neutral
excitation energies of the system. These poles lie just below the real axis in the complex
frequency plane (in the continuous part of the spectrum these are smeared out to a branch
cut).  Therefore the function $\chi^R (z)$ of the complex variable $z$ is an analytic
function in the upper half plane $\Im z >0$. This analytic property is important in
diagrammatic perturbation theory as the density response is an essential ingredient in the
calculation of the screened interaction $W$ as is, for instance, used in the $GW$
approximation. A violation of the analytic properties of $\chi^{R}$ would lead to
incorrect analytic properties of the Green's function. It is therefore a relevant question
to ask whether any approximate expression for the polarizability $\mP^R (\w)$ gives the
correct analytic properties of $\chi^R (\w)$. We show here that this is the case whenever
$\mP^{R}$ is PSD and the interaction matrix $v$ is PSD as well (repulsive interaction).

From the Dyson equation~\eqref{dyson} for the response function we can write the retarded
component in terms of the polarizability as
%---------------------------%
\be
\chi^R (\w) = \mP^R (\w)\left( 1-v\mP^R(\w) \right)^{-1}.
\ee
%---------------------------%
For our proof it turns out to be advantageous to look at the function 
$v^{1/2}\chi^R v^{1/2}$ which can be written as
\be
v^{1/2}\chi^R v^{1/2}= \frac{v^{1/2}\mP^{R}(\w)v^{1/2}}{1-v^{1/2}\mP^R(\w) v^{1/2}}
\ee
where we used the PSD property of $v$ to take the square root operation. The
advantage of this expression is that $v^{\half} \mP^R(\w) v^{\half} $ is PSD whereas
$v \mP^R(\w)$ does not need to be PSD. 
Clearly since $v^{1/2}$ is frequency independent it does not change the analytic 
properties of $\chi^R(\w)$.

We want to know whether $\chi^R (z)$ has a pole in upper half of the 
complex plane, i.e., for $z=x+iy$ with $y >0$. 
This is either the case when $\mP^R (z)$ has a pole at $z$ or when
$1-v^{1/2} \mP^R (z)v^{1/2}$ has a zero. From Eq.~(\ref{PR_rep}) we have
\be
\mP^R (x+iy) = \int \frac{d\omega'}{2 \pi} \frac{\tilde{\mB} (\w')}{x-\w' + i y}
\label{PR_repz}
\ee
and therefore
\bea
\Im \mP^R (z) &=& - y  \int \frac{d\omega'}{2 \pi} \frac{\tilde{\mB} (\w')}{(x-\w')^2 + y^2 },  \label{imPR} \\
\Re \mP^R (z) &=&  \int \frac{d\omega'}{2 \pi} \frac{(x-\w') \tilde{\mB} (\w')}{(x-\w')^2 + y^2 }, \label{rePR}
\eea
%---------------------------%
where we used the short-hand notation $\Im \mP^R = (\mP^R-\mP^A)/2i $ and $\Re \mP^R =
(\mP^R+\mP^A)/2$.  Since
$y>0$ and $\tilde{\mB}$ is an integrable function both integrals are finite and $\mP^R
(z)$ does not have a pole. It remains to consider the possibility that the operator
$1-v^{1/2}\mP^R (z)v^{1/2}$ has a zero eigenvalue.  If this was the case then there would exist an
eigenvector $| \lambda \rangle$ for which $\left( 1-v^{1/2}\mP^R(z)v^{1/2} \right) \left| \lambda
\right\ket = 0$, and hence
\begin{subequations}
\bea
\label{re_condition}
\left\bra \lambda \right| 1-v^{1/2} \Re\mP^R(z)v^{1/2} \left| \lambda 
\right\ket &=& 0,\\
\label{im_condition}
\left\bra \lambda \right| v^{1/2} \Im \mP^R(z)v^{1/2} \left| \lambda 
\right\ket &=& 0.
\eea
\end{subequations}
Using Eq.~(\ref{imPR}) we rewrite Eq. (\ref{im_condition}) as 
%---------------------------%
\bea
\label{Pr}
&&\langle \lambda |v^{1/2} \Im \mP^R(x+iy)v^{1/2} | \lambda \rangle\nn\\
&&=-y \int_{0}^{\infty}\frac{d\w'}{2\pi} \langle \lambda | v^{1/2}
\tilde{\mB}(\w') v^{1/2} | \lambda \rangle \ell(\w'),
\eea
%---------------------------%
where we took into account that $\tilde{\mB}(\w) = -\tilde{\mB}(-\w)$ 
to write the integral between $0$ and $\infty$,
and defined the function
$\ell(\w)$ according to
%---------------------------%
\be
\ell(\w) = \left [  \frac{1}{(x-\w)^2 + y^2}-\frac{1}{(x+\w)^2 + y^2} \right].
\ee
%---------------------------%
The function $\ell(\w)$ is odd and positive (negative) for positive frequencies and for
positive (negative) $x$ values.  Therefore $\ell(\w)$ has a definite sign on the positive
frequency axis when $x$ is nonzero. Let us first consider this case, i.e., $x \neq 0$.
Since we assumed that $y>0$ and since $ \langle \lambda | v^{1/2} \tilde{\mB}(\w') v^{1/2}| \lambda
\rangle\geq 0$ for positive frequencies the only way to have
$\left\bra \lambda \right| v^{1/2} \Im \mP^R(z) v^{1/2}\left| \lambda \right\ket = 0$ is to demand that
$\langle \lambda | v^{1/2} \tilde{\mB}(\w')v^{1/2} | \lambda \rangle=0$ for all $\w'$.  This however,
would imply from Eq.~\eqref{rePR} that also $\langle \lambda |v^{1/2} \Re \mP^R(z)v^{1/2} | \lambda
\rangle=0$ which in turn would imply that $\langle \lambda |1-v^{1/2} \mP^R (z)v^{1/2} | \lambda
\rangle=1$ in contradiction with the assumption.  We conclude that $\chi^R (z)$ cannot
have poles in the upper half plane when $x=\Re z \neq 0$. This leaves us with the case
$x=0$.  For $x=0$ the function $\ell(\omega)=0$ and Eq.~\eqref{im_condition} is
automatically satisfied. Instead Eq.~\eqref{re_condition} reads
%---------------------------%
\bea
&&\langle \lambda | 1-v^{1/2}\Re \mP^R(z)v^{1/2} | \lambda \rangle  \nn\\
&&= 1
+2\!\int_{0}^{\infty}\!\! \frac{d\w'}{2\pi}\,
\frac{\langle \lambda | v^{1/2} \tilde{\mB}(\w')v^{1/2} |\lambda \rangle \w'}{ \w'^2 + y^2} \geq 1
 \label{cond3}
\eea
%---------------------------%
since $v^{1/2} \tilde{\mB}v^{1/2}$ is PSD for positive frequencies. In this case too
Eq.~\eqref{re_condition} cannot be satisfied and therefore $\chi^R$ cannot have poles in
the upper half plane when $\mP^{R}$ and $v$ are PSD.

We mention that the situation is different for negative semidefinite interactions. In this case
a similar formula can be defined but with a minus sign in front of the integral.  In fact, for a
strong enough attraction the right hand side of Eq.~\eqref{cond3} can be zero for a
well-chosen value of $y$ and consequently $\chi^R$ can have poles at $z=iy$ in the upper
half plane. The violation of the analytic property occurs for instance in the attractive
Hubbard dimer~\cite{olsen_static_2014}.  As a final remark we note that in a similar 
fashion one can prove that when the spectral function of the self-energy
is PSD then the Green's function is analytic in the upper half of the complex frequency
plane.

%--------------------------------------------------------------
\paragraph*{Implications to the f-Sum Rule.}
%--------------------------------------------------------------
The connection between positivity and the analytic structure has 
important consequences for
the {\em f-sum rule}, which relates the first momentum of the retarded density response
function to the equilibrium density $n_0(\bx)$
%-------------------------%
\be
\label{fsum_rule}
\int\!\! \w \chi^R(\bx,\bx'\w) \,d\w  = -i\pi\grad'\cdot\grad
\left(n_0(\bx')\delta(\br-\br')\right)\delta_{\sigma\sigma'}.
\ee
%-------------------------%
The derivation of the $f$-sum rule assumes that
the integral over the retarded response function can be closed on the upper 
half-plane\cite{stefanucci_nonequilibrium_2013}, i.e.,
$\chi^R$ is analytic in this region. In accordance with the results 
of this Section the analytic assumption is verified 
for those approximations which fulfill the PSD property.

%*****************************************************************************************

%*****************************************************************************************
\section{PSD diagrammatic expansion}
%*****************************************************************************************
\label{psd-sec}

\subsection{Formulation with noninteracting Green's functions}
\label{psd-rules-sec}

\begin{figure}[t!]
\centering
       \includegraphics[width=0.9\linewidth]{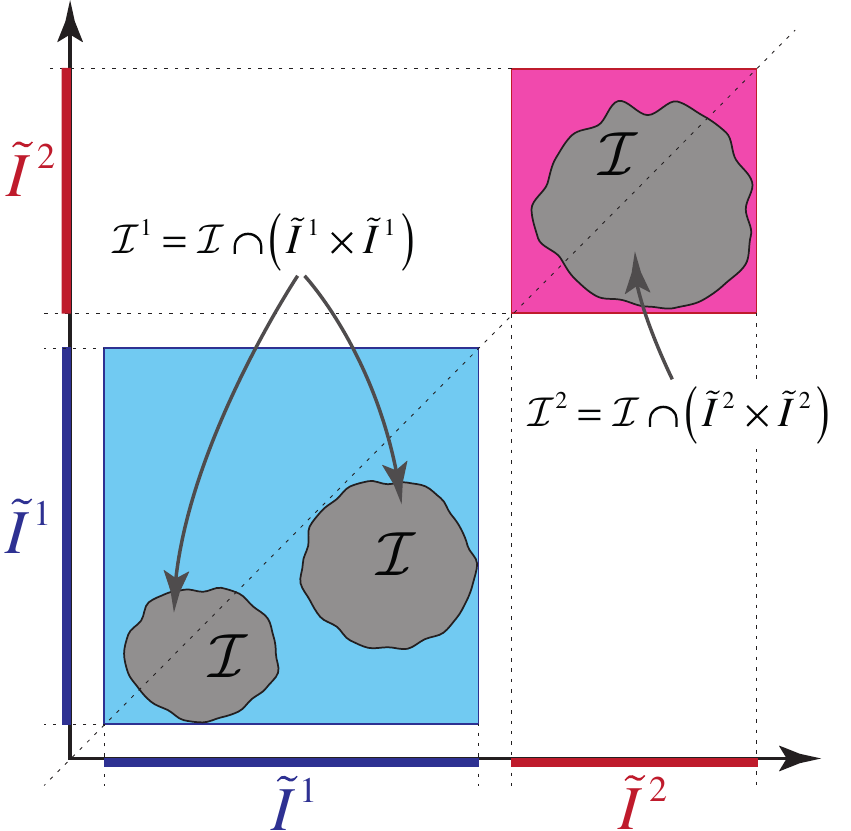}
\caption{(Color online) Decomposition $\bigcup_{\a}\mathcal{I}_{N}^{\a}=\mathcal{I}_{N}$
  of the $\mathcal{I}_N$ subset (denoted as three disjoint gray areas) into a union of
  (two in this example) product sets.
 \label{fig:venn}}
\end{figure}

In Section \ref{theory} we have given a diagrammatic proof of the PSD property of the
exact polarizability.  We further showed that this PSD property is essential to guarantee
the correct analytic structure of the density response function. In practice, however, the
polarizability is calculated in a MBPT fashion by considering a subclass of Feynman
diagrams and hence the PSD property is not, in general, a built-in property of the
approximation.  The proof given in Section \ref{posp} paves the way for a diagrammatic theory of PSD polarizabilities
which is alternative to the more standard MBPT. We have seen that the PSD property follows
from the formation of perfect squares and that these squares are the sum of partitions. In
contrast MBPT gives a sum of diagrams where each diagram is a sum of partitions but, in
general, do not form perfect squares. The most general MBPT approximation to the
polarizability when written in terms of partitions (or equivalently in terms of products
of half-diagrams) reads
%---------------------------%
\bea
\label{p_mbpt}
i\mP^<(1,2) &=& \sum_{N=1}^{\infty} \sum_{(j_1,j_2)\in \mathcal{I}_N}
\sum_{\substack{P\in 
\pi^{(j_{1}j_{2})}_{N,p} \\ Q \in \pi^{(j_{1}j_{2})}_{N,q}} }(-)^{P+Q} \nn\\
&\times& \sum_{\up\uq} D_{\up\uq}^{(j_2)}(2)D^{(j_1)^*}_{P(\up)Q(\uq)}(1),
\eea
%---------------------------%
where $\mathcal{I}_N$ is a subset of the product set $I_{N}\times I_{N}$ and for any given
couple $(j_{1},j_{2})$ the sums over $P$ and $Q$ run over a subset
$\pi^{(j_{1}j_{2})}_{N,p}$ and $ \pi^{(j_{1}j_{2})}_{N,q}$ of the permutation group
$\pi_{N}$. The minimal number of additional partitions to add in order to turn the MBPT
approximation into a PSD approximation is found by imposing on $\mP^{<}$ the mathematical
structure in Eq.~\eqref{pwithpart}. This is achieved as follows, see 
also Fig.~\ref{fig:venn}. Let
$\{\tilde{I}_{N}^{\a}\}$ be a set of disjoint subsets of $I_{N}$ with the property that
the union of the product sets
%---------------------------%
\be
\bigcup_{\a}\tilde{I}_{N}^{\a}\times\tilde{I}_{N}^{\a}\supset\mathcal{I}_N
\label{unionia}
\ee
%---------------------------%
and contains the least number of elements of $I_{N}\times I_{N}$.  Since the subsets
$\tilde{I}_{N}^{\a}$ are disjoint the sets $\mathcal{I}_{N}^{\a}=\mathcal{I}_N \cap
(\tilde{I}_{N}^{\a}\times\tilde{I}_{N}^{\a})$ are disjoint too and due to
Eq.~\eqref{unionia} we have that $\bigcup_{\a}\mathcal{I}_{N}^{\a}=\mathcal{I}_{N}$. For
any given $\a$ we then consider the smallest subgroups of the permutation group $\pi_{N}$
with the property that
%---------------------------%
\bea
\tilde{\pi}_{N,p}^{\a} \supset \bigcup_{(j_1 j_2 )\in 
\mathcal{I}_{N}^{\a}} \pi_{N,p}^{(j_1,j_2)},
\\
\tilde{\pi}_{N,q}^{\a} \supset \bigcup_{(j_1 j_2 )\in 
\mathcal{I}_{N}^{\a}} \pi_{N,q}^{(j_1,j_2)}.
\eea
%---------------------------%
%---------------------------%
By construction the polarizability  
\bea
\label{p_psd}
i\mP_{\rm PSD}^<(1,2) &=& \sum_{N=1}^{\infty} \sum_{\a}\sum_{j_1,j_2\in 
\tilde{I}_N^{\a}}
\sum_{\substack{P\in 
\tilde{\pi}^{\a}_{N,p} \\ Q \in \tilde{\pi}^{\a}_{N,q}} }
(-)^{P+Q} \nn\\
&& \sum_{\up\uq} D_{\up\uq}^{(j_2)}(2)D^{(j_1)^*}_{P(\up)Q(\uq)}(1)
\eea
%---------------------------%
contains all partitions of Eq.~\eqref{p_mbpt} plus the minimal number of additional
partitions to form perfect squares. Consequently the polarizability in Eq.~\eqref{p_psd} is
a diagrammatic PSD approximation.  More precisely we can say that a PSD diagrammatic
approximation to $\mP$ is not the sum of MBPT diagrams, rather it is the sum of partial or
decorated diagrams (the partitions) with internal vertices either on the minus or the plus
branch of the Keldysh contour. Examples of the PSD procedure are 
presented in Section \ref{example}.

As a final remark we mention that another cutting procedure based on time-ordered half
diagrams has been used by Sangalli {\em et al.}~\cite{sangalli_double_2011} for the
Bethe-Salpeter kernel of the so-called second RPA approximation.  In that work the Lehmann
product structure was important to satisfy an identity for the determinant of the
Bethe-Salpeter kernel which guarantees that no spurious poles occur in the density
response function obtained by solving the Bethe-Salpeter equation.  However, the procedure
proposed in Ref.~\onlinecite{sangalli_double_2011} cannot be used to obtain a direct
expression for the spectral function, thus making it difficult to address the PSD property.
The difficulty has its origin in the fact that the standard time-ordered formalism is not
the natural formalism to express the spectral function.  As it has been shown in this work
the Keldysh contour technique facilitates enormously the calculation of the spectral
function of any diagram, the result being a product of time-ordered and anti-time-ordered
half diagrams joined by lesser and greater Green's function lines.

%*****************************************************************************************
\subsection{Formulation with dressed Green's functions}
%*****************************************************************************************
In the previous Section we discussed how to generate PSD diagrammatic approximations for
the polarizability. A diagrammatic approximation is the sum of partitions and for the PSD
property to be satisfied the product of half-diagrams resulting from the cut partitions
has to form the sum of perfect squares. The possibility of cutting a partition relies on
Eq.~\eqref{cutting}, according to which the product of two $g^{\lessgtr}$ yields a single
$g^{\lessgtr}$. This property, however, is valid only for noninteracting Green's
functions.  It would be extremely useful to formulate a cutting rule for partitions
written in terms of dressed Green's functions. The main advantage of a dressed PSD
formulation is the absence of polarizability-diagrams with self-energy insertions, thus
enabling us to work exclusively with skeleton diagrams.

Here we consider partitions with dressed Green's function lines and show how to write
these partitions as the product of two half-diagrams.  We therefore need to replace
Eq.~\eqref{cutting} with some other equation where $G^{\lessgtr}$ is expressed as the
``product'' of two functions.  The main idea has been discussed in detail in our recent
work on the self-energy.~\cite{stefanucci_diagrammatic_2014} In frequency space the
greater and lesser Green's function read
%---------------------------%
\be
G^{\gtrless} (\bx_1t_1,\bx_2t_2) =  \mp i\int \frac{d\w}{2\pi} A^{\gtrless}(\bx_1,\bx_2;\w)e^{-i\w(t_1-t_2)},
\ee
%---------------------------%
where $A^<(\w) \equiv f(\w )A(\w)$ is the removal part of the spectral function $A(\w)$
whereas $A^>(\w) \equiv (1-f(\w ))A(\w)$ is the addition part of $A(\w)$, and $f(\w)$ is
the zero temperature Fermi function.  If the self-energy is PSD then both $A^>$ and $A^<$
are PSD. We expand the matrix $A^{\gtrless}(\w)$ in terms of its eigenvalues
$a_n^{\gtrless}(\w)\geq 0$ and eigenvectors $u_n(\w,\bx)$
%---------------------------%
\bea
A^{\gtrless}(\bx_1,\bx_2;\w) = \sum_n a_n^{\gtrless}(\w) u_n(\w,\bx_1)u_n^*(\w,\bx_2)
\eea
%---------------------------%
and  define the square root matrix  as
%---------------------------%
\be
\sqrt{A^{\gtrless}}(\bx_1,\bx_2;\w) = \sum_n \sqrt{a_n^{\gtrless}(\w)} u_n(\w,\bx_1)u_n^*(\w,\bx_2).
\ee
%---------------------------%
We then make the rule that when cutting a partition the internal lines are
$G^{\minus\minus}$ for the left half, $G^{\plus\plus}$ for the right half, and
%---------------------------%
\be
\sqrt{G^{\gtrless}} (\bx_1t_1,\bx_2t_2) =  \int \frac{d\w}{2\pi} \sqrt{A^{\gtrless}}(\bx_1,\bx_2;\w)e^{-i\w(t_1-t_2)}
\ee
%---------------------------%
for the dangling lines of the two halves. For the reverse operation of gluing the two
halves we make the rule that the ``product'' of two dangling lines is defined according to
\begin{subequations}
\bea
\int\! d\by dt\, 
\sqrt{G^<}_{\bx_1\by}(t_1,t)\sqrt{G^<}_{\by\bx_2}(t,t_2) 
\label{cutting_dressed}&=&iG^<(1,2) ,\quad\\
\int\! d\by dt\, \sqrt{G^>}_{\bx_1\by}(t_1,t)\sqrt{G^>}_{\by\bx_2}(t,t_2)
&=&-iG^>(1,2).\quad\quad
\eea
\label{dresscut}
\end{subequations}
These equations replace Eq.~\eqref{cutting} and the analogous for $g^{>}$ in the dressed
case. Except that for an additional time-integration Eqs.~\eqref{dresscut} have the same
matrix-product structure as in the undressed case and it is straightforward to verify that
the gluing of two half-diagrams gives back the original partition. This implies that if a
certain sum of partitions with undressed Green's functions $g$ is PSD, and hence it can be
written as in Eq.~\eqref{p_psd}, then the same sum of partitions with dressed Green's
functions $G$ is PSD too since the corresponding polarizability can be written as
%---------------------------%
\bea
\label{dressed_p_psd}
i\mP_{\rm PSD}^<(1,2) &=&\sum_{N=1}^{\infty} \sum_{\a}\sum_{j_1,j_2\in \tilde{I}_N^{\a}}
\sum_{\substack{P\in \tilde{\pi}^{\a}_{N,p} \\ Q \in \tilde{\pi}^{\a}_{N,q}} } (-)^{P+Q} \nn\\
&\times& \int \!d1_{p}\ldots dN_{p}\int\! d1_{q}\ldots dN_{q} \nn\\
&\times&  D_{1_{p}\ldots N_{p},1_{q}\ldots N_{q}}^{(j_2)}(2)\,
D^{(j_1)^*}_{P(1_{p}\ldots N_{p})Q(1_{q}\ldots N_{q})}(1),\quad\quad
\eea
%---------------------------%
where we introduced the short-hand notation $n_{p}=(p_{n},t^{(p)}_{n})$ ($n=1,\ldots,N$)
and $m_{q}=(q_{m},t^{(q)}_{m})$ ($m=1,\ldots,N$) as well as
%---------------------------%
\be
\int dn_{p}=\sum_{p_{n}}\int dt^{(p)}_{n}
\;;\quad
\int dm_{q}=\sum_{q_{m}}\int dt^{(q)}_{m}.
\ee
%---------------------------%
In Eq.~\eqref{dressed_p_psd} the $D$'s represent the half diagrams with dangling lines
$\sqrt{G^\lessgtr}$ and external vertices in $ 1_{p}\ldots N_{p},1_{q}\ldots N_{q}$.

%*****************************************************************************************
\section{Examples}
\label{example}
%*****************************************************************************************
In this section we consider some commonly used diagrammatic approximations to the
polarizability and address the PSD property for each of them.

\begin{figure}[t!]
\centering
       \includegraphics[width=\linewidth]{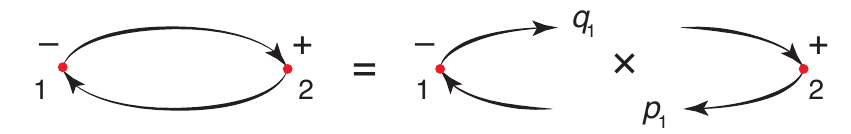}
\caption{(Color online) Partition and decomposition in half-diagrams of the RPA bubble
  diagram.}
\label{example1}
\end{figure}
\paragraph{Zeroth order} In Fig.~\ref{example1} we show the RPA bubble. This 
diagram can be partitioned in only one way and the decomposition in terms of half-diagrams
is shown on the right of the equality sign.  According to the cutting rules for dressed
diagrams we have
\bea
D_{1_{p} 1_{q}}^{*}(1) &=&\sqrt{G^<}_{\bx_1 
p_1}(t_1,t^{(p)}_{1})\sqrt{G^>}_{q_1\bx_1}(t^{(q)}_{1},t_1),\nn\\
D_{1_{p} 1_{q}}(2)&=& 
\sqrt{G^<}_{p_{1}\bx_2}(t^{(p)}_{1},t_2)\sqrt{G^>}_{\bx_2 
q_{1}}(t_2,t^{(q)}_{1}),\nn
\eea
and it is straightforward to verify that the RPA bubble can be written as
%---------------------------%
\bea
\label{P1_lesser}
\mP^{<}_{\textrm{RPA}}(1,2) &=&- i\int\, d1_{p} d1_{q}\;
D_{1_{p}1_{q}}(2)\, D_{1_{p}1_{q}}^{\ast}(1).
\eea
%---------------------------%
The RPA $\mP$ has clearly the structure of Eq. (\ref{dressed_p_psd}) and therefore the
spectrum for the response function is positive.

%---------------------------%
\begin{figure}[b!]
\centering
       \includegraphics[width=\linewidth]{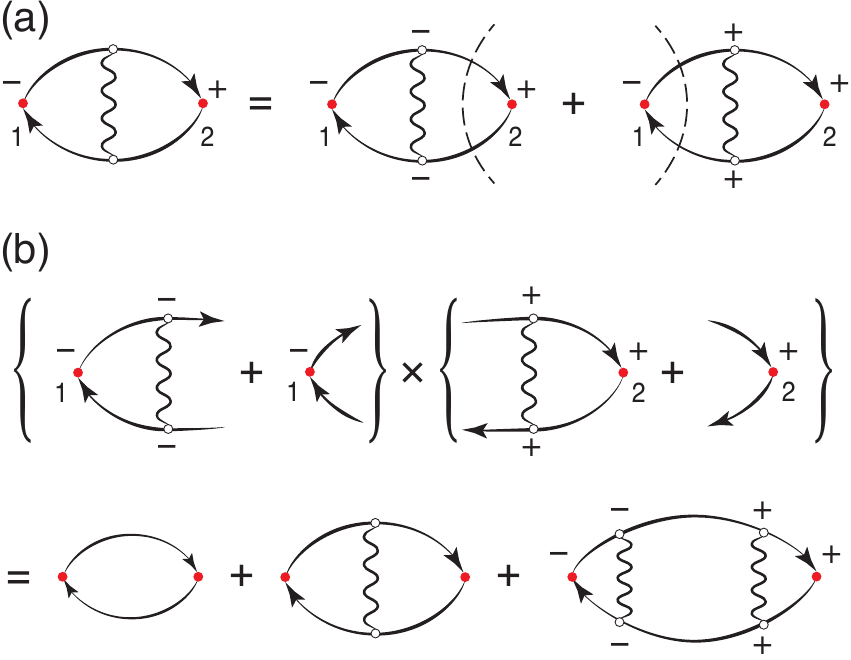}
\caption{(Color online) (a) Partition and decomposition in half-diagrams of the vertex
  diagram. The wiggly line denotes the bare interaction $v$. (b) The minimal set of
  additional partitions to restore the PSD property.}
\label{example2}
\end{figure}
%---------------------------%
\paragraph{Simplest vertex} 
In Fig.~\ref{example2}(a) we consider the lowest order (in bare Coulomb interaction)
vertex diagram for $\mP$. The cutting rules yield only two partitions (on the right of the
equality sign) since the bare interaction is local in time and hence $v^{\lessgtr}=0$.
From the decomposition in half-diagrams we see that the vertex diagram does not have the
structure of Eq.~\eqref{dressed_p_psd}. Therefore the PSD property is not guaranteed. The
minimal set of partitions to add follows directly from the rules of Section
\ref{psd-rules-sec}.  The result is shown in Fig.~\ref{example2}(b). By multipling the
half-diagrams in the curly brackets we get two additional diagrams: the RPA bubble and a
partition of the second-order ladder diagram. Here and in the following we use the
convention that an internal vertex without $-/+$ labels implies a summation over $-/+$.
From this example we also infer that the sum of only the RPA bubble and the vertex diagram
is not, in general, PSD.  Noteworthy this sum constitute the so called {\em exact
  exchange} (EXX) approximation to the kernel of Time-Dependent Density Functional Theory
(TDDFT). The EXX kernel has been calculated in
Ref.~\onlinecite{hellgren_exact-exchange_2009} for the case of closed shell atoms and it
was found that it has poles in the upper half of the complex frequency plane. This
incorrect analytic behavior is a direct consequence of the relation between the PSD and
the analytic properties, as discussed in Section~\ref{analytic-sec}. The correct analytic
properties of the EXX kernel can be restored by adding the partition of the second-order
ladder diagram shown in Fig.~\ref{example2}(b).

\begin{figure}[t!]
\centering
     \includegraphics[width=\linewidth]{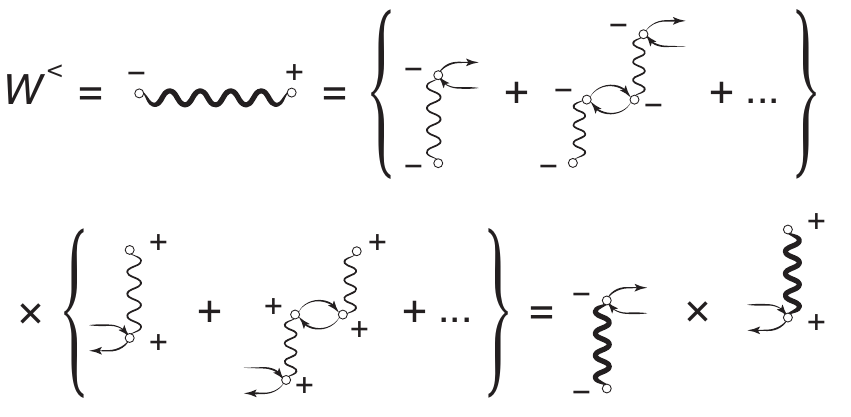}
\caption{Decomposition of the screened interaction (thick wiggly line) $W^{<}$ in
  half-diagrams. The decomposition of $W^{>}$ is analogous and given by the bottom diagram
  with $-\leftrightarrow +$.}
\label{example3}
\end{figure}

\paragraph{RPA screening} 
In bulk systems the electron screening plays a crucial role and, therefore, one usually
works with interaction-skeletonic diagrams in which the bare interaction $v$ is replaced
by some screened interaction $W$. Below we apply the theory developed in
Section~\ref{psd-rules-sec} to a few diagrammatic approximations in which $W$ is the RPA
screened interaction, i.e., $W=v+v\mP_{0}W$ where $\mP_{0}=-iGG$.  Taking into account that
partitions with isolated $+/-$ islands do not contribute to the polarizability we
immediately conclude that $W^{--}$ and $W^{++}$ should not be cut along the internal
Green's function lines. For the lesser/greater $W$-lines we note that
$W^{<}=W^{--}\mP_{0}^{<}W^{++}$ and $W^{>}=W^{++}\mP_{0}^{>}W^{--}$, see
Fig.~\ref{example3}. Therefore the cut of a $W^{\lessgtr}$ amounts to a cut of the two
Green's function lines in $\mP_{0}^{\lessgtr}$.
%---------------------------%
%---------------------------%
%---------------------------%
\begin{figure}[b!]
\centering
       \includegraphics[width=\linewidth]{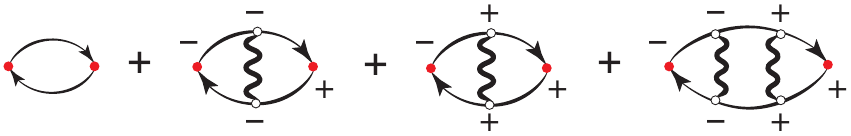}
\caption{(Color online) The simplest PSD approximation to the polarizability in 
terms of the screened interaction $W$ (thick wiggly line).}
\label{example4}
\end{figure}
%---------------------------%

The simplest diagrammatic approximation in terms of $W$ is given by the four partitions of
Fig.~\ref{example2}(b) in which $v\to W$, see Fig.~\ref{example4}.  We observe that the
sum of them does not give the full vertex diagram since the partitions with $W^{-+}=W^{<}$
and $W^{+-}=W^{>}$ are missing. These partitions would vanish if, 
instead of the RPA $W$, we used an externally given {\em static} $W$ 
like, e.g., a Yukawa-type interaction.
We will come back to this example in Section \ref{numsec}
where we give explicit expressions of 
the diagrams of Fig.~\ref{example4} for the electron gas and show the
crucial role played by the second-order ladder diagram for the PSD property.

%---------------------------%
\begin{figure}[t!]
\centering
       \includegraphics[width=\linewidth]{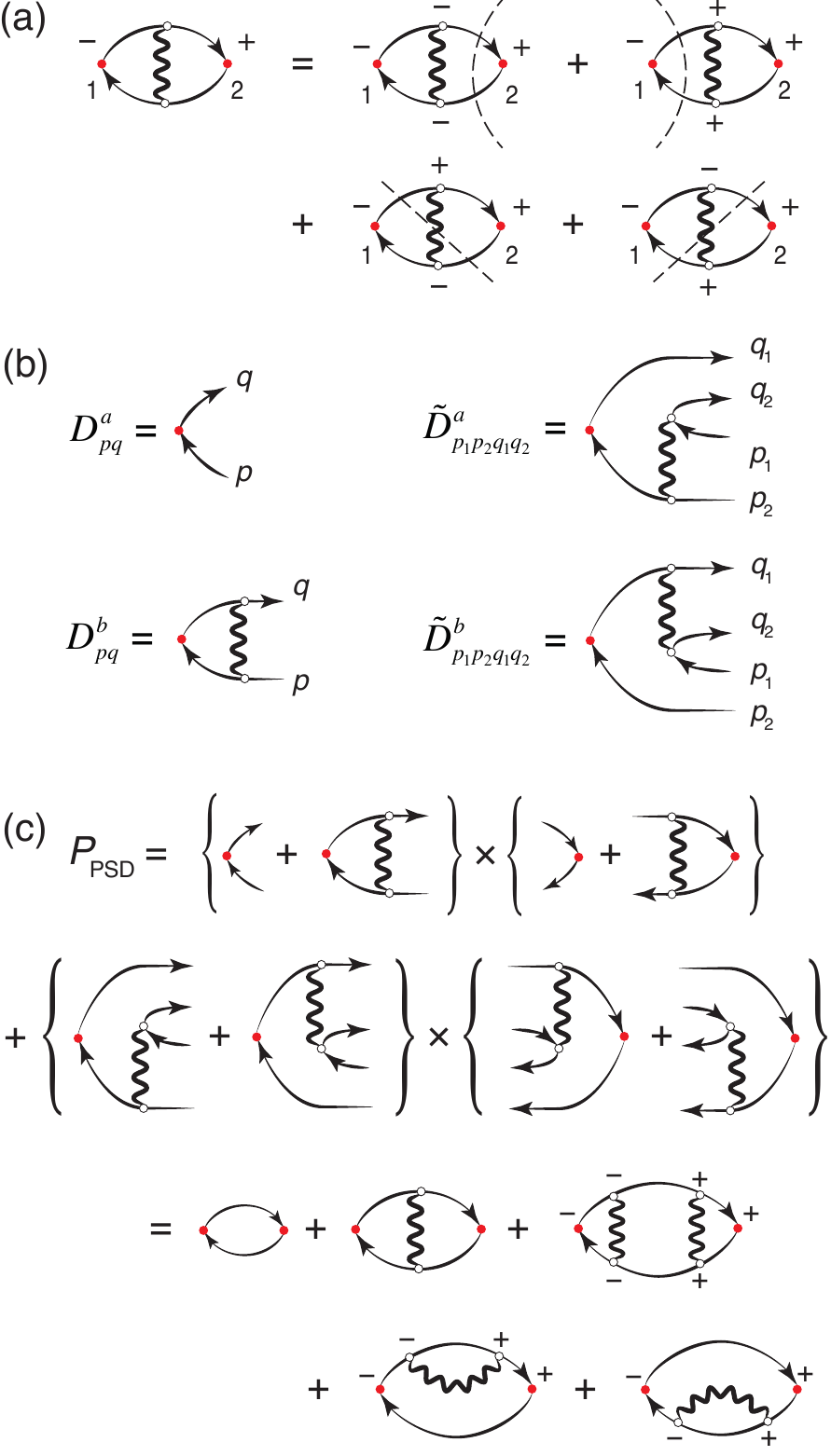}
\caption{(Color online) (a) Decomposition of the vertex diagram with screened interaction
  line into half-diagrams. (b) Constituent half-diagrams. (c) The resulting PSD polarizability.}
\label{example5}
\end{figure}
%---------------------------%
\paragraph{First order}
The decomposition into half-diagrams of the full first order (in screened Coulomb
interaction) vertex diagram is shown in Fig.~\ref{example5}(a).  Since the times of the
internal vertices can be different (nonlocal interaction) we have four different
partitions. In two of them we only cut the $G$-lines while in the other two we also cut
the $W$-line (the cut of the $W$-line leads to half-diagrams with four dangling lines).
Thus, we have two half-diagrams with one particle-hole ($D^{(a)}$ and $D^{(b)}$) and two
half-diagrams with two particle holes ($\tilde{D}^{(a)}$ and $\tilde{D}^{(b)}$). Naming
each half-diagram as shown in the figure~\ref{example5}(b) we can write (omitting the
integrals over the vertices to be glued as well as dependence on the external vertices)
%---------------------------%
\be
i\mP^{<}=D^{(a)}D^{(b)^{\ast}}+D^{(b)}D^{(a)^{\ast}}
+\tilde{D}^{(a)}\tilde{D}^{(b)^{\ast}}+\tilde{D}^{(b)}\tilde{D}^{(a)^{\ast}}, 
\ee 
%---------------------------%
which does not have the structure of Eq. (\ref{dressed_p_psd}) and hence it is not PSD.
Applying the rules of Section \ref{psd-rules-sec} we find the minimal set of partitions to
add in order to restore the PSD property
%---------------------------%
\be 
i\mP^{<}_{\rm PSD}=\sum_{ij=a,b}D^{(i)}D^{(j)^{\ast}}+
\sum_{ij=a,b}\tilde{D}^{(i)}\tilde{D}^{(j)^{\ast}},
\ee 
%---------------------------%
where in each of the sums the indices $i$ and $j$ independently take values $a$, $b$.
The resulting diagrammatic approximation to $\mP$ is illustrated in Fig.~\ref{example5}(c).
The important message of this example is that the additional diagrams are not necessarily
skeletonic in $G$ (occurrence of self-energy insertions, viz. the last 
two diagrams in the
figure). Thus attention has to be paid when restoring the PSD property using a
\emph{dressed} $G$. In order to avoid double countings one should not dress the $G$ with
the same self-energy appearing in the diagrams of the PSD polarizability.

%---------------------------%
\begin{figure}[t!]
\centering
       \includegraphics[width=\linewidth]{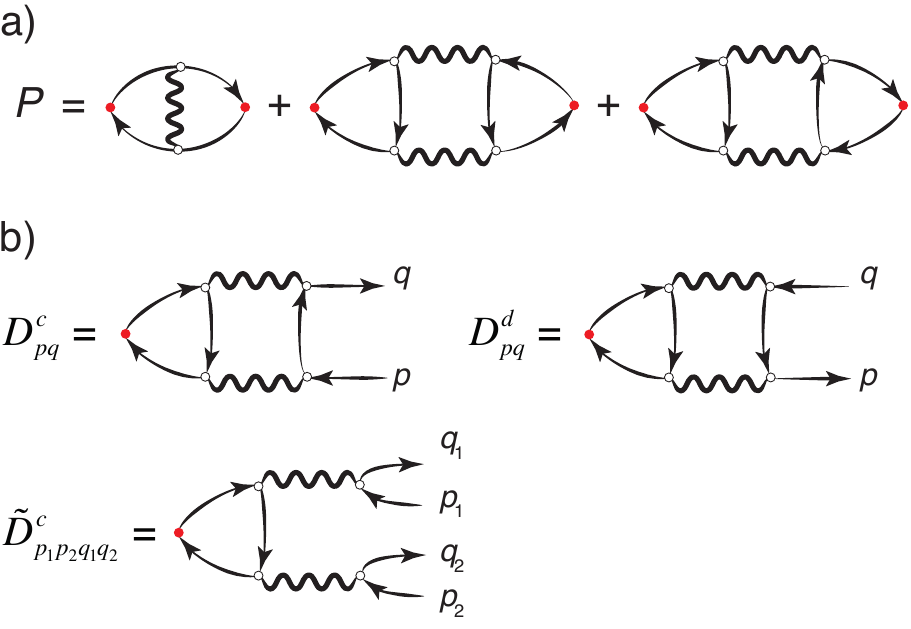}
\caption{(Color online) (a) Diagrammatic approximation to the polarizability. (b)  
Constituent half-diagrams in addition to those presented in Fig.~\ref{example5}(b).}
\label{example6}
\end{figure}
\begin{figure}[t!]
\centering
       \includegraphics[width=\linewidth]{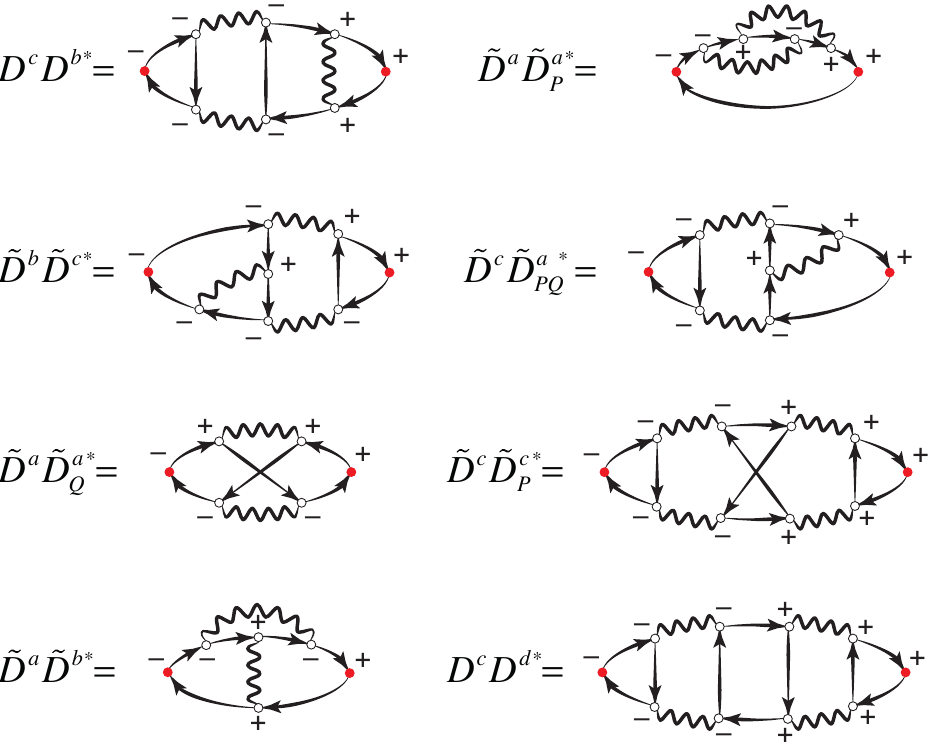}
\caption{(Color online) A few additional partitions of Eq.~\eqref{psdlastexmp}.}
\label{example7}
\end{figure}

%---------------------------%
\paragraph{GW exchange-correlation kernel}
We conclude this section with another important example. In
Ref.~\onlinecite{von_barth_conserving_2005} it was shown how to generate conserving
approximations to the TDDFT kernel using a variational principle {\em 
\`a la}
Luttinger-Ward.~\cite{luttinger_ground-state_1960} The underlying variational functional of
Luttinger-Ward is a functional of the bare interaction $v$ and the Green's function
$G$. To lowest order in $v$ one can show that the variational principle leads to the EXX
approximation. It is possible to extend the Luttinger-Ward idea to functionals of the
screend interaction $W$ and the Green's function $G$.\cite{almbladh_variational_1999} In
this case the lowest order approximation is the ``time-dependent $GW$'' (TD$GW$)
approximation; and the TDDFT kernel (more precisely its convolution with two $\mP_{0}$'s) is
given by the sum of the diagrams in Fig.~\ref{example6}(a) (see also Sec.~III.B of
Ref.~\onlinecite{von_barth_conserving_2005}). These are the same diagrams evaluated by
Sternemann {\em et al.}~\cite{sternemann_correlation-induced_2005} and by Huotari {\em et
  al.}~\cite{huotari_electron-density_2008} for the electron gas in order to explain the
double-plasmon shoulder in the absorption spectrum of sodium. By partitioning each diagram
of the approximate polarizability we find that $\mP^{<}$ can be written in terms of four
half-diagrams with one particle-hole and three half-diagrams with two particle-holes. Some
of them have already been introduced in Fig.~\ref{example5}(b), the new ones are shown in
Fig.~\ref{example6}(b) and the expression of $\mP^{<}$ in terms of them is (again omitting
integrals and the dependence on the external vertices)
%---------------------------%
\bea
i\mP^{<}&=&\tilde{D}^{(a)}\left(\tilde{D}^{(a)}_{PQ}+\tilde{D}^{(b)}+\tilde{D}^{(b)}_{PQ}
+\tilde{D}^{(c)}_{P}+\tilde{D}^{(c)}_{Q}\right)^{\ast}
\nn\\&+&\tilde{D}^{(b)}\left(\tilde{D}^{(b)}_{PQ}+\tilde{D}^{(a)}+\tilde{D}^{(a)}_{PQ}
+\tilde{D}^{(c)}_{P}+\tilde{D}^{(c)}_{Q}\right)^{\ast}
\nn\\&+&\tilde{D}^{(c)}\left(\tilde{D}^{(c)}+ \tilde{D}^{(c)}_{PQ}+
\tilde{D}^{(a)}_{P}+ \tilde{D}^{(a)}_{Q}+\tilde{D}^{(b)}_{P}+\tilde{D}^{(b)}_{Q}\right)^{\ast}
\nn\\&+&D^{(a)}\left(D^{(b)}+D^{(c)}+D^{(d)}\right)^{\ast}
\nn\\&+&\left(D^{(b)}+D^{(c)}+D^{(d)}\right)D^{(a)^{\ast}}, 
\label{eq:TDGW}
\eea
%---------------------------%
where the half-diagrams with subindex $P$ (and/or $Q$) are calculated at permuted values
$2_{p},1_{p}$ (and/or $2_{q},1_{q}$). For instance, $\tilde{D}^{(c)}
\tilde{D}^{(c)^\ast}_{PQ}$ is a short form of the integral
\[
\int \!d1_{p}d2_{p}\int \!d1_{q}d2_{q}\; \tilde{D}^{(c)}_{1_p2_p1_q2_q}(2)\, \tilde{D}^{(c)^\ast}_{2_p1_p2_q1_q}(1).
\]
The polarizability of Eq.~(\ref{eq:TDGW}) is not, in general, PSD since it does not have the
form of Eq.~\eqref{dressed_p_psd}. The minimal addition to restore the PSD property
follows from the rules of Section~\ref{psd-rules-sec} which give
%---------------------------%
\be
i\mP^{<}_{\rm PSD}=\sum_{ij=a,b,c,d}D^{(i)}D^{(j)^{\ast}}+ \sum_{ij=a,b,c}\sum_{P,Q\in\pi_{2}}
\tilde{D}^{(i)}\tilde{D}^{(j)^{\ast}}_{PQ}.
\label{psdlastexmp}
\ee
%---------------------------%
The first sum leads to $4\times4=16$ partitions whereas the second sum leads to
$3\times(2^2\times3)=36$ partitions. In Fig.~\ref{example7} some 
representative ones are shown.
%---------------------------%
%---------------------------%

%*****************************************************************************************
\section{On the Positivity of the Bethe-Salpeter kernel}
\label{bsesec}
%*****************************************************************************************
%-------------------------%
\begin{figure}[b!]
\centering
  \includegraphics[width=\linewidth]{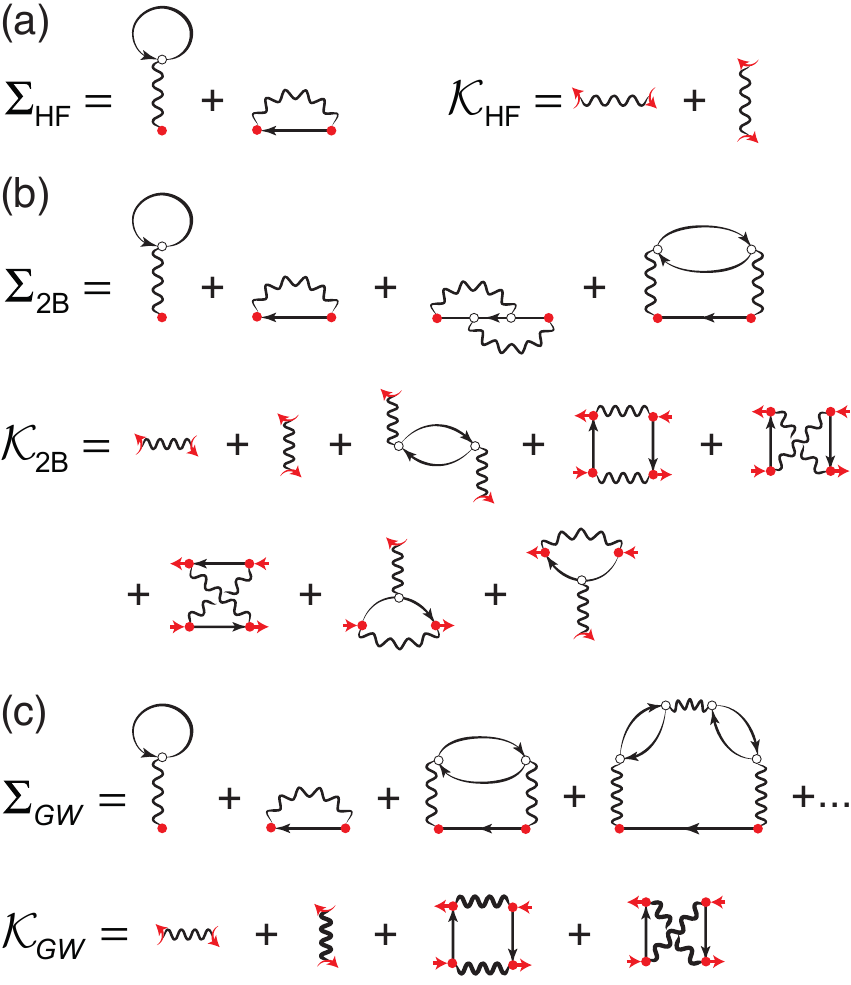}
\caption{(Color online) Self-energy $\S$ and the corresponding BSE 
kernel for (a) Hartree-Fock (b) Second order Born and (c) $GW$ 
approximations.}
\label{fig:bse-kernel}
\end{figure}
%---------------------------%
So far we have studied only approximations to the irreducible response
function consisting of a finite number of diagrams. 
However, typically approximations beyond the RPA involve an infinite series
of diagrams conveniently resummed through the Bethe-Salpeter equation
(BSE)~\cite{salpeter_relativistic_1951,strinati_application_1988}.  A natural question to
ask is whether the corresponding spectral function is PSD.  To answer we have to find
the diagrammatic structure encoded in the kernel of the
BSE. The BSE is a Dyson-like equation for the four-point reducible polarizability $\mL
(12;34)$ and it is obtained as the response to a non-local scalar 
potential  $u(4,3)$~\cite{strinati_application_1988}
%---------------------------%
\bea
\label{bse}
\mL(12;34)&=&\! - \frac{\delta G(1,2) }{ \delta u(4,3)}=\mL_0(12;34)\nn\\
+&&\!\!\!\!\!\!\!\!\int\! d(5678) 
G(1,5)G(7,3)\mK(56;78)\mL(82;64)\quad
\eea
%---------------------------%
where $\mL_0(12;34) = G(1,4)G(2,3)$ and the four-point reducible 
kernel $\mK$ is given by
$\mK(12;34) = -i \delta\Sigma(1,3) / \delta G(4,2)$.
The variation of the Green's function is related to the two-particle Green's
function and therefore $\mL$ is related to the two-particle 
excitation spectrum.  By taking the limit
$3\to1^+$ and $4\to2^+$ we obtain an equation for the response 
function since $\chi(1,2) = i
\mL(12;1^+2^+)$.

\begin{figure}[b!]
\centering
       \includegraphics[width=\linewidth]{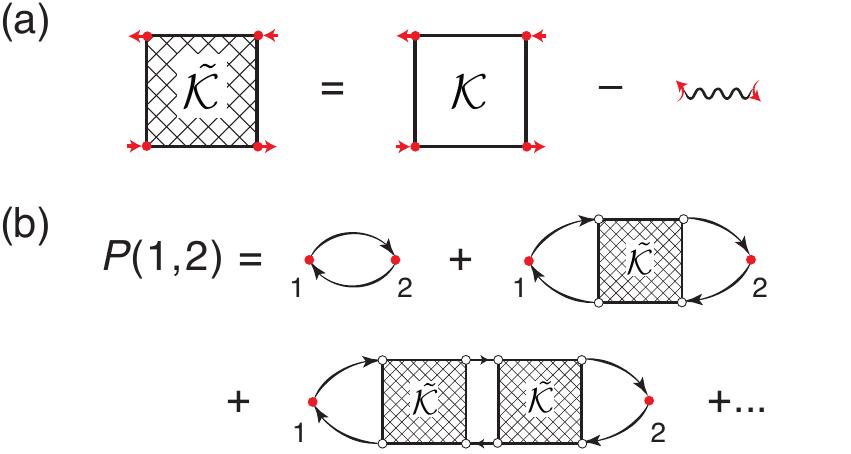}
\caption{(Color online) (a) Two-particle irreducible and one 
interaction line irreducible
  kernel $\tilde{\mK}$.  (b) The diagrammatic expression for 
  the polarizability in terms of BSE kernel $\tilde{\mK}$.}
\label{fig:kernel_bubble}
\end{figure}

\begin{figure}[t!]
\centering
       \includegraphics[width=\linewidth]{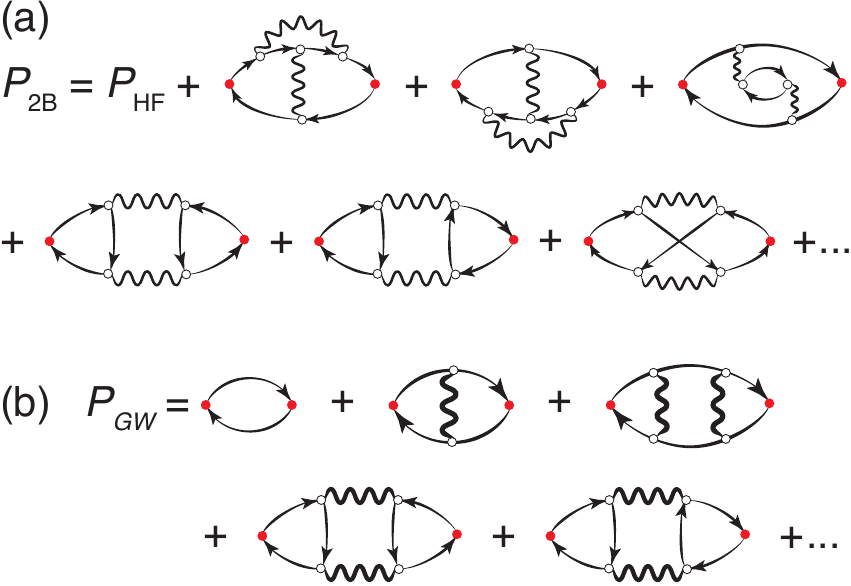}
\caption{(Color online) Approximations for the polarizability by using BSE kernel with
  various self-energy approximations. (a) 2B polarizability. (b) $GW$ 
  polarizability.}
\label{fig:p_bse}
\end{figure}

\begin{figure}[h!]
\centering
       \includegraphics[width=\linewidth]{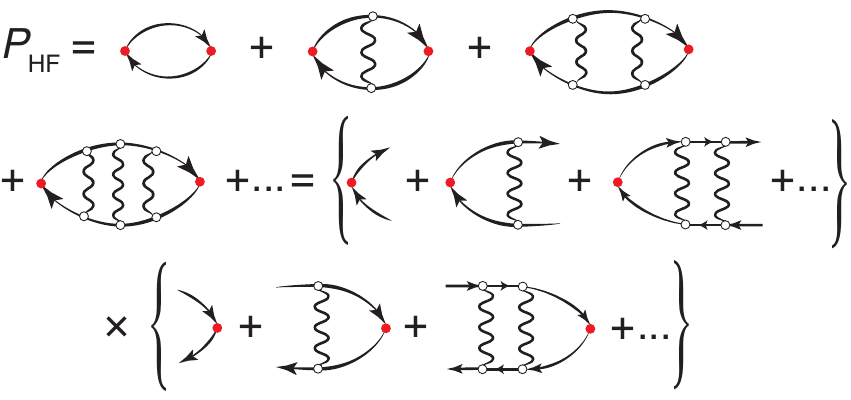}
\caption{(Color online) The polarizability calculated from the HF BSE kernel is PSD. }
\label{fig:hf_bse}
\end{figure}

From standard approximations to the self-energy, e.g., Hartree-Fock (HF), second order Born (2B)
or the $GW$ approximation, we can derive a diagrammatic expression 
for the kernel $\mK$,
see Fig.~\ref{fig:bse-kernel}.  By defining a two-particle irreducible and one
interaction line irreducible kernel $\tilde{\mK}$ as (see also 
Fig.~\ref{fig:kernel_bubble}(a))
\bea 
\tilde{\mK} (12;34) = \mK(12;34)
-i\delta(1,3)\delta(2,4) v(1,2), 
\eea 
we can write the polarizability  as an infinite series of response diagrams
\bea
\label{p_bse}
\mP(1,2) = \mP_0(1,2)+i\!\int \!d(3456)\,G(5,1)G(1,3)\tilde{\mK}(34;56)\nn\\
\times G(4,2)G(2,6)+\ldots.\quad
\eea
%-------------------------%
%---------------------------%

\noindent
where $\mP_0(1,2)=-iG(1,2)G(2,1)$.
The diagrammatic expression for this equation is shown in Fig.~\ref{fig:kernel_bubble}(b).
By using the kernels in Fig.~\ref{fig:bse-kernel} we obtain
the approximations shown in Fig.~\ref{fig:p_bse}. The HF approximation for the BSE kernel
yields the diagrams shown in the Fig.~\ref{fig:hf_bse}, which we can easily see to be 
PSD. From this example we also deduce that 
if a kernel cannot be partitioned then the corresponding 
polarizability is PSD.
A  commonly
used approximation to study the exitonic properties of solids is the {\em static} $GW$
approximation~\cite{strinati_application_1988,sangalli_double_2011,olsen_static_2014}
with kernel
%---------------------------%
\be
\tilde{\mK}^{(0)}_{\textrm{GW}}(12;34)= i\delta(1,2)\delta(3,4) 
W(1,3),
\ee
where the functional derivative of $W$ with respect to $G$ is 
neglected.
%-------------------------%
%---------------------------%
\noindent
This approximation leads to the polarizability of
Fig.~\ref{fig:hf_bse} in which the bare interaction lines are 
replaced by statically screened ones.  Therefore, the {\em
  static} $GW$ approximation yields PSD spectra. For the full $GW$ approximation some
of the half-diagrams have been already worked out in 
Fig.~\ref{example6} and the resulting diagrams
after the gluing procedure have been shown in Fig.~\ref{example7}. From this figure we see
that the PSD procedure leads to new types of diagrams which are not 
obtained via the
iteration of the BSE. Thus, we conclude that the BSE polarizability with $GW$ kernel does not necessarily
have PSD spectra. 
The same is true for the 2B approximation as can be seen from the half-diagrams generated by 
the last diagram on the first line of  Fig.~\ref{fig:p_bse}. By applying the PSD
procedure we will generate diagrams which are not obtained via iteration of the
BSE. Therefore, even the 2B approximation is not a PSD approximation 
for the BSE.  These simple
examples show that the kernels generated by conserving 
$\Phi$-derivable self-energies 
\cite{baym_conservation_1961,baym_self-consistent_1962,luttinger_ground-state_1960} 
do not need to be PSD.
%-------------------------%
%---------------------------%

%*****************************************************************************************
\section{Numerical results}
\label{numsec}
%*****************************************************************************************
As an illustration of our method we compute the spectral functions $\tilde{\mB}(k,\omega)$ of the
polarizability $\mP(k,\omega)$ for the three-dimensional homogeneous electron
gas. For convenience we introduce here the spectral functions for the positive
$\tilde{\mB}^>(k,\omega)$ and negative $\tilde{\mB}^<(k,\omega)$ frequencies and scale them by the factor
of $8\pi\alpha r_s$ in order to make the zeroth order (the Lindhard polarization function)
density independent
\be
\tilde{\mB}^{(0)}(k,\omega)=\left\{
\begin{array}{lr}\frac{\omega}{k}&\frac{|\omega|}{k}\le 2-k,\\
\frac{1}{k}-\frac{1}{4k}\big(k-\frac{\omega}{k}\big)^2&\;\;|2-k|<\frac{|\omega|}{k}\le 2+k,
\end{array}
\right.
\label{eq:0}
\ee
where we expressed $k$ in units of the Fermi momentum $k_F=1/(\alpha r_s)$ and $\omega$ in
units of the Fermi energy $\epsilon_F=k_F^2/2$. Here $\alpha=[4/(9\pi)]^{1/3}$ and $r_s$
is the standard measure of the system's density--the Wigner-Seitz radius--expressed in
units of the Bohr radius.

%---------------------------%
\begin{figure}[b!]
\centering
\includegraphics[width=\columnwidth]{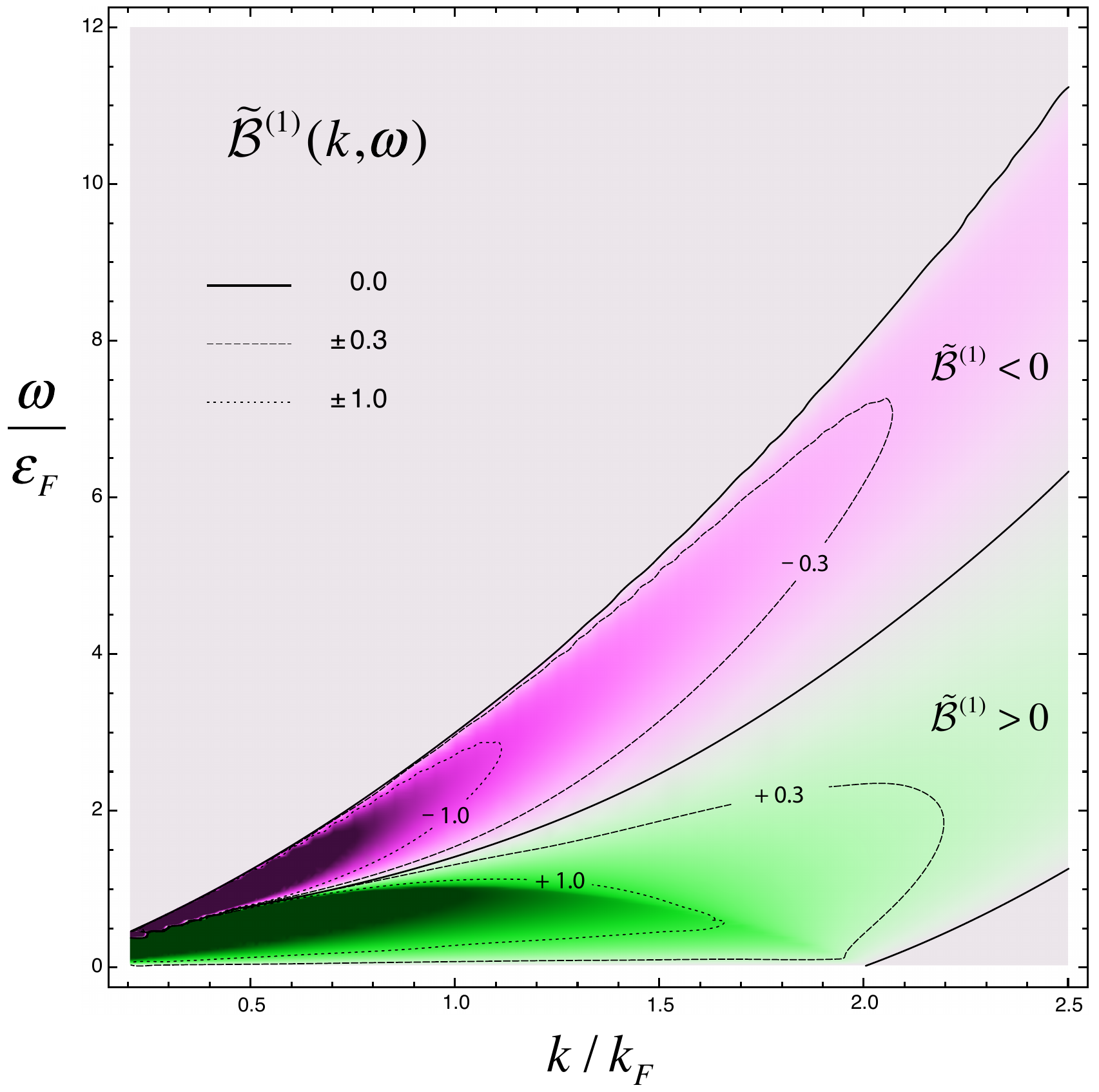}
\caption{(Color online) Distribution of positive (green) and negative (pink) values of the
  first order spectral function of the vertex diagram in 
  Fig.~\ref{example2}(b) in the $k-\omega$ plane.
  The lines represent isospectral curves with values $\pm 0.3$  
  (dashed) $\pm 0.1$ (dotted) and $0$ (solid).
  \label{fig:chi_dens_plot}}
\end{figure}
%---------------------------%

In 1958 J. Hubbard diagrammatically studied the correlation energy of a free-electron
gas~\cite{hubbard_description_1958} and introduced what is now known as the \emph{local
  field factor} ($f(k,\omega)$, same notation as in the original manuscript is used). A
very interesting introduction to the historical development of this concept and its
importance for the density functional theory can be found in
Ref.~\onlinecite{giuliani_quantum_2005}. In essence, it provides a simple way to go
beyond RPA both in the treatment of the density response function and the total
energy of a many-body system and relates between the exact and zeroth order
polarizabilities $v(k)f(k,\omega)=[\mP(k,\omega)]^{-1}-[\mP^{(0)}(k,\omega)]^{-1}$. Thus,
every advancement in the calculation of the proper density response function leads
to our improved knowledge of the local field factor and, correspondingly, of the density
functionals. After Hubbard's original static approximation $f(k,\omega)\approx
k^2/(k^2+k_F^2)$ there were numerous works to compute the local field factor using the
above relation, notably the exact long and short wavelength limits (see
Ref.~\onlinecite{giuliani_quantum_2005} and references therein). Diagrammatically, the
simplest case of the first order diagrams (in terms of the bare Coulomb interaction) was
computed in a concise form by Engel and Vosko~\cite{engel_wave-vector_1990} for the static
case. Importantly, they considered two types of diagrams, the proper first order response
and the first order self-energy insertions and demonstrated a rather large cancellations
between these two contributions. In a full generality the frequency dependent first-order
results were obtained by Holas, Aravind, and 
Singwi~\cite{holas_dynamic_1979}. However,
the analytic form is rather complicated and can only be expressed in 
terms of a one-dimensional integral. We will use these results for a comparison and, therefore, numerical
details concerning the evaluation of this integral are presented in
Appendix~\ref{ApB}. In fact, already the first order vertex diagram 
of Fig.~\ref{example2}(b) demonstrates the problem of
standard MBPT. In Fig.~\ref{fig:chi_dens_plot} we depict its momentum and energy
resolved spectral function $\tilde{\mB}^{(1)}$ computed according to the expression of
Ref.~\onlinecite{holas_dynamic_1979}. The pink shaded area denotes a part of the
particle-hole continuum where the 1st order spectral function is negative.

%---------------------------%
\begin{figure}[t!]
\centering
\includegraphics[width=\columnwidth]{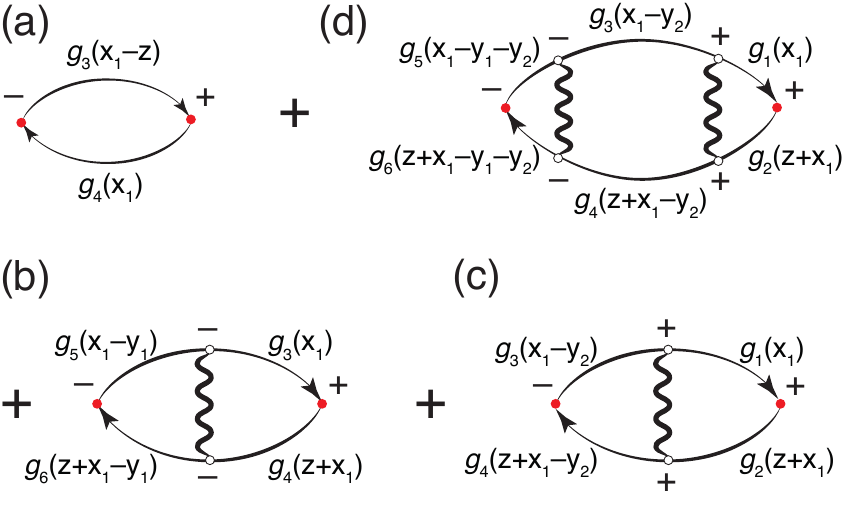}
\caption{(Color online) Sum of these zero to second order polarizability diagrams yields
  positive spectral function. Lines with arrows denote the electron propagator
  $G_0(k,\omega)$, whereas wavy-lines stand for bare or screened Coulomb
  interaction. Vertices are labeled with $+$ ($-$) if they belong to the positive
  (negative) time ordering part of the Keldysh contour.
 \label{fig:p012}}
\end{figure}
%---------------------------%

In order to solve this problem we use our method and consider the irreducible
polarizability diagrams shown in Fig.~\ref{example2}(b) and Fig.~\ref{example4}: a
particular second order diagram must be added in order to compensate for the negative sign
of $\tilde{\mB}^{(1)}$. The two sets of diagrams are topologically identical, however, the first
one is given in terms of bare Coulomb, while the second contains screened interacting
lines. To be general we will start with the second more complicated case and derive the
first case by making the limit $w(k)\rightarrow\Omega(k)\rightarrow\infty$ in the plasmon
pole approximation for the screened Coulomb interaction: 
%---------------------------%
\be
W_{0}^{--}(k,\omega)=\frac{v(k)}2\!
\left[\frac{w(k)}{\omega-\Omega(k)+i\eta}-\frac{w(k)}{\omega+\Omega(k)-i\eta}\right].
\label{eq:W0}
\ee
%---------------------------%
In this expression $w(k)=t(k)\Omega^2(0)/\Omega(k)$, $0\le t(k)\le 1$ is the plasmonic
spectral weight, and $\Omega(k)$ is the plasmonic dispersion with $\Omega(0)=4\sqrt{\alpha
  r_s/(3\pi)} \epsilon_F$, where $\epsilon_F$ is the Fermi energy.
For the numerical integration we define the bare time-ordered Green's 
function as\cite{stefanucci_diagrammatic_2014}
\begin{equation}
G_0^{--}(k,\omega)=\frac{B(k)}{\omega-\epsilon_k-i\eta}+\frac{A(k)}
{\omega-\epsilon_k+i\eta}.
\label{eq:g0}
\end{equation}
In non-interacting systems $B(k)=n_k$ and $A(k)=1-n_k$ with $n_k$ denoting the occupation
of the state with momentum $k$. The frequency integrations can be done completely
analytically (facilitated by the {\sc mathematica} computer algebra system (CAS)) whereas
for the remaining momentum integrations one has to rely on
numerics~\cite{pavlyukh_time_2013}. The starting point are the four diagrams depicted at
Fig.~\ref{fig:p012}. The momentum flows are explicitly shown. There are in general many
possibilities to assign momenta to propagators.  Our choice is dictated by the matter of
convenience and is not unique. In order to further simplify notations we adopt the
following short forms: $A_i\equiv A(x_i)$, $B_i\equiv B(x_i)$, $C_i\equiv
\frac12v(y_i)w(y_i)$, and introduce the function
\[
\mathcal{H}_i(a,\Omega)=\frac{A_i}{a-\Omega-\epsilon_i}+\frac{B_i}{a+\Omega-\epsilon_i}. 
\]
We also recall that $\tilde{\mB}^{>}(k,|\omega|)=\tilde{\mB}^{<}(k,-|\omega|)$. Therefore, it is
sufficient to consider only one case, e.g., $\omega>0$.
\begin{widetext}
The results of frequency integration are
\begin{subequations}
\label{eq:P012}
\begin{eqnarray}
\tilde{\mB}_{a}^{<}(z,\zeta)\!\!&=&\!\!-\pi \!\!\int\!\!\frac{d^3x_1}{(2\pi)^3}A_3B_4\delta(\zeta-\epsilon_4+\epsilon_3),\\
\tilde{\mB}_{b}^{<}(z,\zeta)\!\!&=&\!\!-\pi 
\!\!\int\!\!\frac{d^3x_1}{(2\pi)^3}\!\int\!\!\frac{d^3y_1}{(2\pi)^3}
\frac{\mathcal{H}_6(\epsilon_4,\Omega_1)-\mathcal{H}_5(\epsilon_3,\Omega_1)}
{\epsilon_3-\epsilon_4-\epsilon_5+\epsilon_6}C_1A_3B_4\delta(\zeta-\epsilon_4+\epsilon_3),\\
\tilde{\mB}_{c}^{<}(z,\zeta)\!\!&=&\!\!-\pi 
\!\!\int\!\!\frac{d^3x_1}{(2\pi)^3}\!\int\!\!\frac{d^3y_2}{(2\pi)^3}
\frac{\mathcal{H}_2(\epsilon_4,\Omega_2)-\mathcal{H}_1(\epsilon_3,\Omega_2)}
{\epsilon_3-\epsilon_4-\epsilon_1+\epsilon_2}C_2A_3B_4\delta(\zeta-\epsilon_4+\epsilon_3),\\
\tilde{\mB}_{d}^{<}(z,\zeta)\!\!&=&\!\!-\pi 
\!\!\int\!\!\frac{d^3x_1}{(2\pi)^3}\!\int\!\!\frac{d^3y_1}{(2\pi)^3}\!\int\!\!\frac{d^3y_2}{(2\pi)^3}
\frac{\mathcal{H}_6(\epsilon_4,\Omega_1)-\mathcal{H}_5(\epsilon_3,\Omega_1)}
{\epsilon_3-\epsilon_4-\epsilon_5+\epsilon_6}
\frac{\mathcal{H}_2(\epsilon_4,\Omega_2)-\mathcal{H}_1(\epsilon_3,\Omega_2)}
{\epsilon_3-\epsilon_4-\epsilon_1+\epsilon_2}
C_1C_2A_3B_4\delta(\zeta-\epsilon_4+\epsilon_3).\;\;
\end{eqnarray}
\end{subequations}

They are quite general and can be used to obtain e.g. plasmonic contribution. If, however,
results for bare Coulomb interacting lines are needed we take the limits and obtain:
\begin{subequations}
\label{eq:p012}
\begin{eqnarray}
\tilde{\mB}_{a}^{<}(z,\zeta)\!\!&=&\!\!-\pi \!\!\int\!\!\frac{d^3x_1}{(2\pi)^3}A_3B_4\delta(\zeta-\epsilon_4+\epsilon_3),\\
\tilde{\mB}_{b}^{<}(z,\zeta)\!\!&=&\!\!-\pi \!\!\int\!\!\frac{d^3x_1}{(2\pi)^3}\int\!\!\frac{d^3y_1}{(2\pi)^3}v(y_1)
\frac{A_6-A_5}{\epsilon_3-\epsilon_4-\epsilon_5+\epsilon_6}A_3B_4\delta(\zeta-\epsilon_4+\epsilon_3),\\
\tilde{\mB}_{c}^{<}(z,\zeta)\!\!&=&\!\!-\pi \!\!\int\!\!\frac{d^3x_1}{(2\pi)^3}\int\!\!\frac{d^3y_2}{(2\pi)^3}v(y_2)
\frac{A_2-A_1}{\epsilon_3-\epsilon_4-\epsilon_1+\epsilon_2}A_3B_4\delta(\zeta-\epsilon_4+\epsilon_3),\\
\tilde{\mB}_{d}^{<}(z,\zeta)\!\!&=&\!\!-\pi \!\!\int\!\!\frac{d^3x_1}{(2\pi)^3}\int\!\!\frac{d^3y_1}{(2\pi)^3}\int\!\!\frac{d^3y_2}{(2\pi)^3}
v(y_1)v(y_2)\frac{A_6-A_5}{\epsilon_3-\epsilon_4-\epsilon_5+\epsilon_6}
\frac{A_2-A_1}{\epsilon_3-\epsilon_4-\epsilon_1+\epsilon_2}
A_3B_4\delta(\zeta-\epsilon_4+\epsilon_3).
\end{eqnarray}
\end{subequations}
\end{widetext}
Notice that the spectral functions in Eqs.~\eqref{eq:P012} and \eqref{eq:p012} are denoted by
the same symbol because their type can always be inferred from the context.  In these
equations $A$, $B$, and $\epsilon$ quantities are labeled by the momenta as shown at
Fig.~\ref{fig:p012}. For instance
\[
B_5\equiv \left\{
\begin{array}{cc}
1&|x_1-y_1-y_2|\le k_F,\\
0&|x_1-y_1-y_2|>k_F,
\end{array}\right.
\]
and $\epsilon_5=(x_1-y_1-y_2)^2/2$.  $\tilde{\mB}_{a}(z,\zeta)\equiv \tilde{\mB}^{(0)}(z,\zeta)$ is
obviously the Lindhard polarization function. $\tilde{\mB}_{b}(z,\zeta)$ and $\tilde{\mB}_{c}(z,\zeta)$
differ only by the permutation of indices and, therefore, are equal in view of the
left-right symmetry of the corresponding diagrams. There are no other topologically
identical diagrams of the first order,
i.e. 
$\tilde{\mB}_{b}(z,\zeta)+\tilde{\mB}_{c}(z,\zeta)=\tilde{\mB}^{(1)}(z,\zeta)$. There are two more partitions
of the second order having the same topology as $\tilde{\mB}_{d}(z,\zeta)$. They have different
combinations of pluses and minuses assigned to the vertices and are 
not considered here, hence $\tilde{\mB}_{d}(z,\zeta)=\tilde{\mB}^{(2)}(z,\zeta)$.

Before analyzing numerical results let us first notice a different proportionality of each
perturbative order to the electron density. Because we scaled all spectral functions such
that $\tilde{\mB}^{(0)}$ is density independent it is easy to see that
$\tilde{\mB}^{(n)}=\mathcal{O}((\alpha r_s)^n)$. Thus, the sum of four terms in
Eqs.~\eqref{eq:p012} is given by a quadratic polynomial in terms of $\alpha r_s$. The
requirement of its positivity leads, therefore, to the following inequality
\begin{equation}
\mathcal{D}=(\tilde{\mB}_{b}^{<}(z,\zeta))^2-\tilde{\mB}_{a}^{<}(z,\zeta)\tilde{\mB}_{d}^{<}(z,\zeta)\le 0,
\label{eq:D}
\end{equation}
which should hold for all values of frequency and momenta for at least one density. This
ensures the positivity of the spectral function at \emph{all} densities. Mathematically,
inequality~\eqref{eq:D} is the Cauchy-Schwarz inequality applied to the integrals of
half-diagrams.

The 1st and 2nd order expressions (Eqs.~\eqref{eq:p012}) as well as the analytic result of
Holas \emph{et al.}~[\onlinecite{holas_dynamic_1979}] (Appendix~\ref{ApB}) suffer from the
logarithmic singularities of the integrated functions. This poses additional challenges for
numerics. We tackle this problem by introducing a small imaginary part $i\delta$ into the
energy denominators. The Monte-Carlo integration is performed as in
Ref.~\onlinecite{stefanucci_diagrammatic_2014} with the use of Mersenne twister 19937
random number generator~\cite{haramoto_efficient_2008}. Additional complication arises due
to the use of bare rather than the plasmonically screened Coulomb interaction, i.e. even
large momentum transfers are possible. In this case we can still map the integration
variables to the $[0,1]$ interval by the logarithmic scaling. For instance, we represent
the vector $\vec y_1$ as follows $r_1=-p\log(x_1)$, $\cos(\theta_1)=-1+2x_2$, $\phi_1=2\pi
x_3$, where $p$ is some suitably chosen constant~\footnote{We set $p=3$. We verified that
  an order of magnitude variation of this parameter has no impact on the accuracy of
  calculations} and $x_i\in [0,1]$. Thus, $d^3y_1=\frac{4\pi p}{x_1}dx_1dx_2dx_3$.

\begin{figure}[]
\centering
\includegraphics[width=\columnwidth]{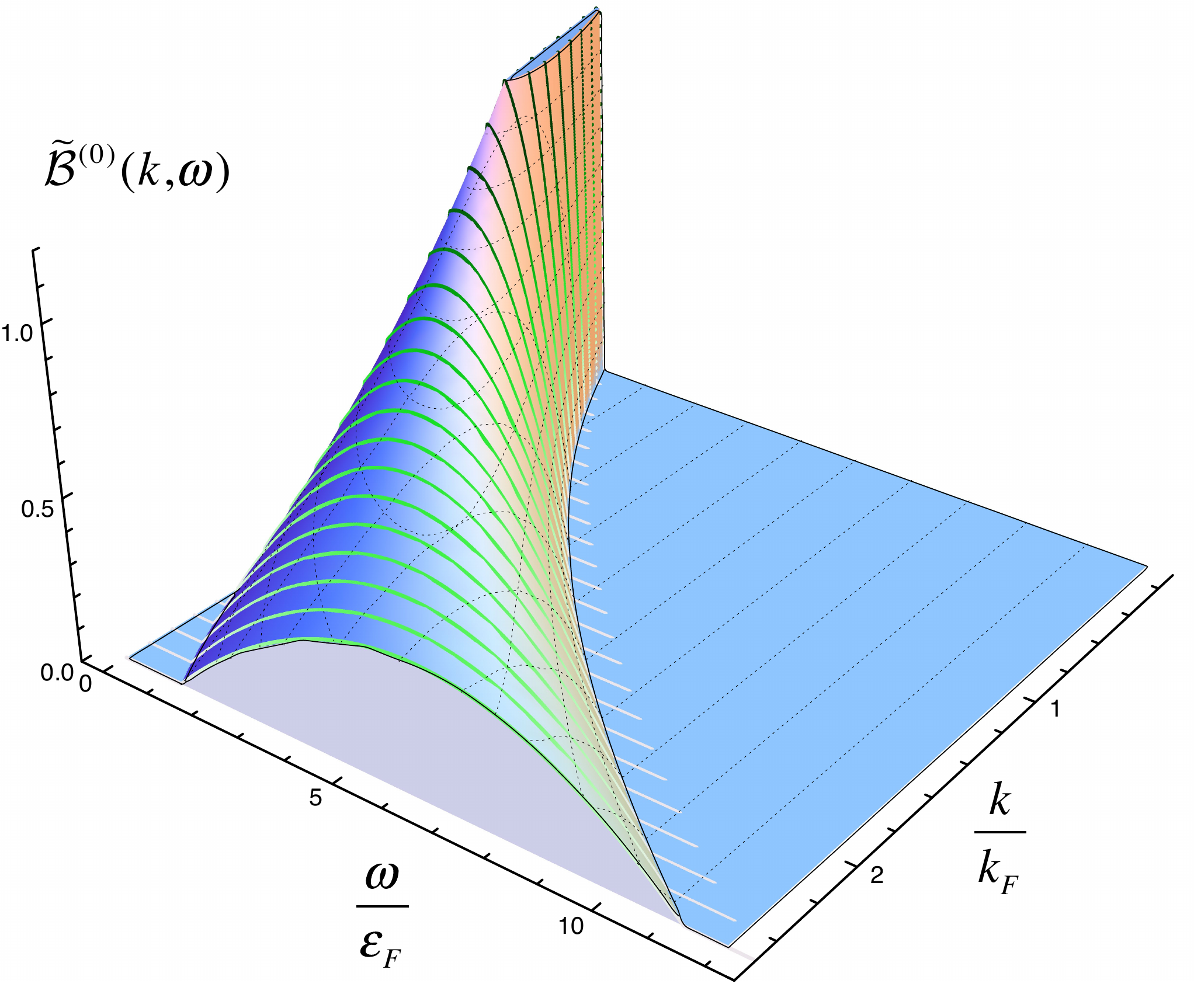}
\includegraphics[width=\columnwidth]{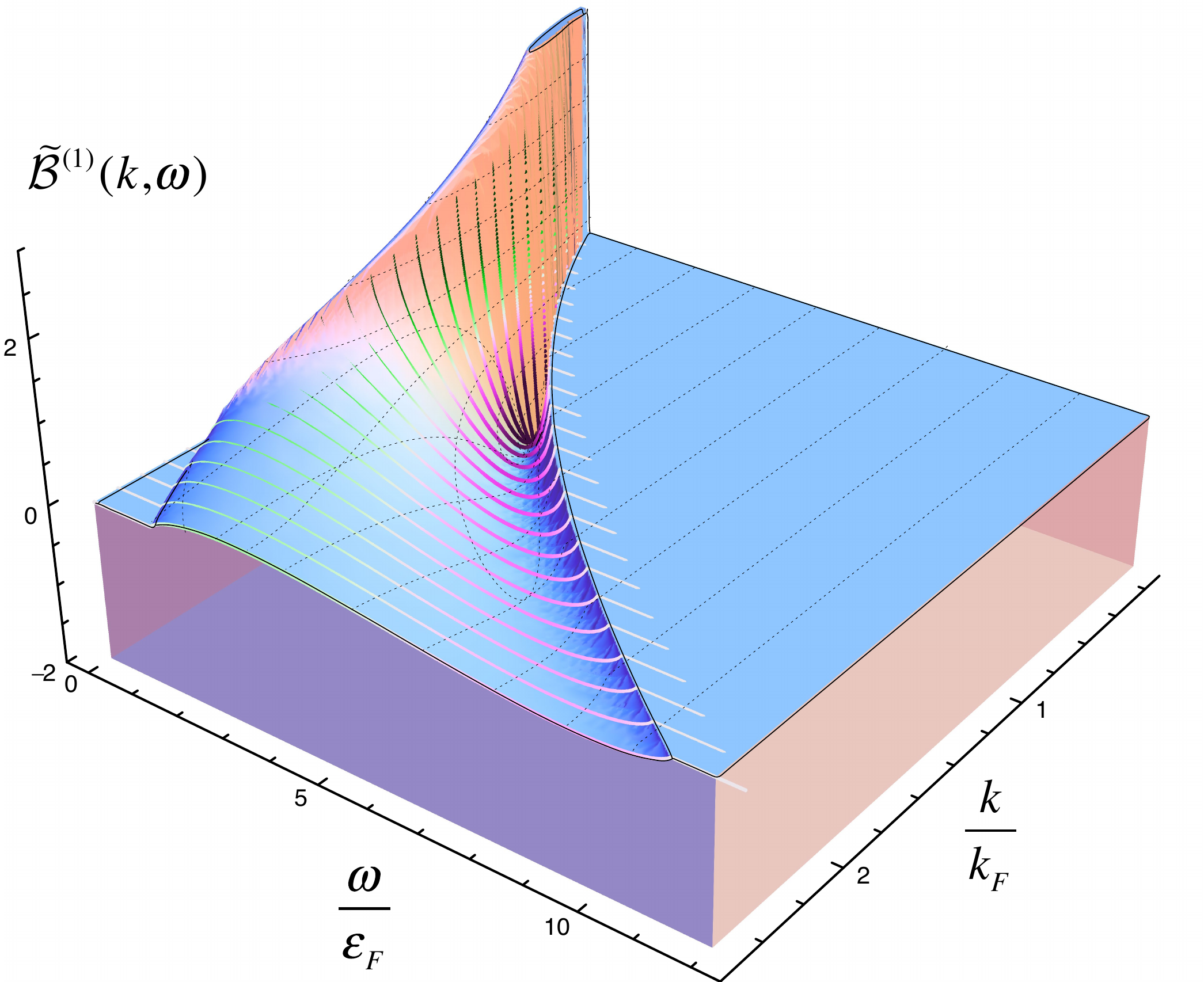}
\includegraphics[width=\columnwidth]{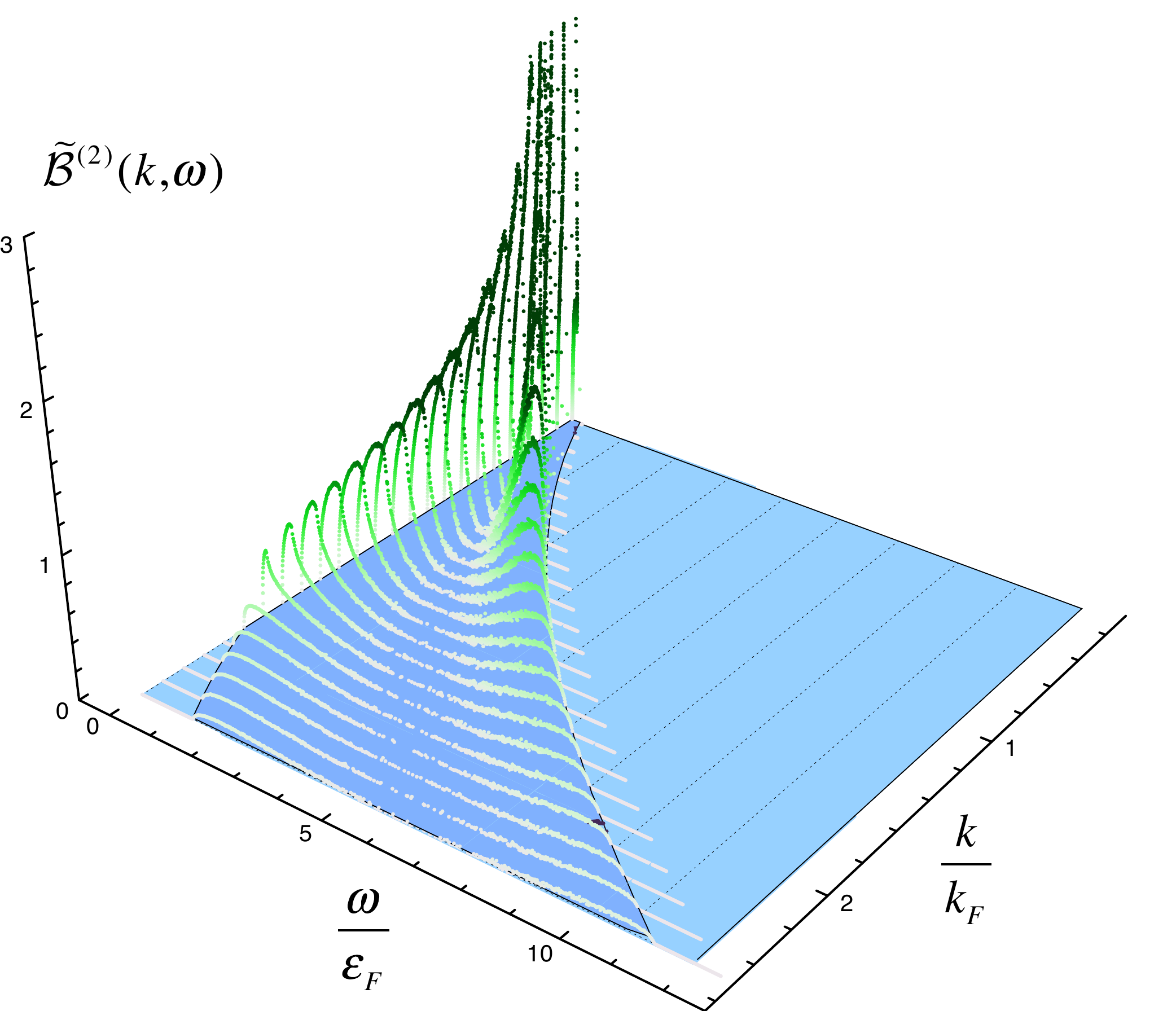}
\caption{(Color online) The scaled spectral function for the polarizability
  $\mP$ of homogeneous electron gas at $r_s=3$. Surfaces denote exact analytical results:
  0th order is given by the Lindhard function (Eq.~\eqref{eq:0}) and 1st order is computed
  according to 1d integral representation of Holas \emph{et
    al.}\cite{holas_dynamic_1979}. Dots denote numerical results obtained by the
  Monte-Carlo calculation. \label{fig:chi}}
\end{figure}

\begin{figure}[t!]
\centering \includegraphics[width=\columnwidth]{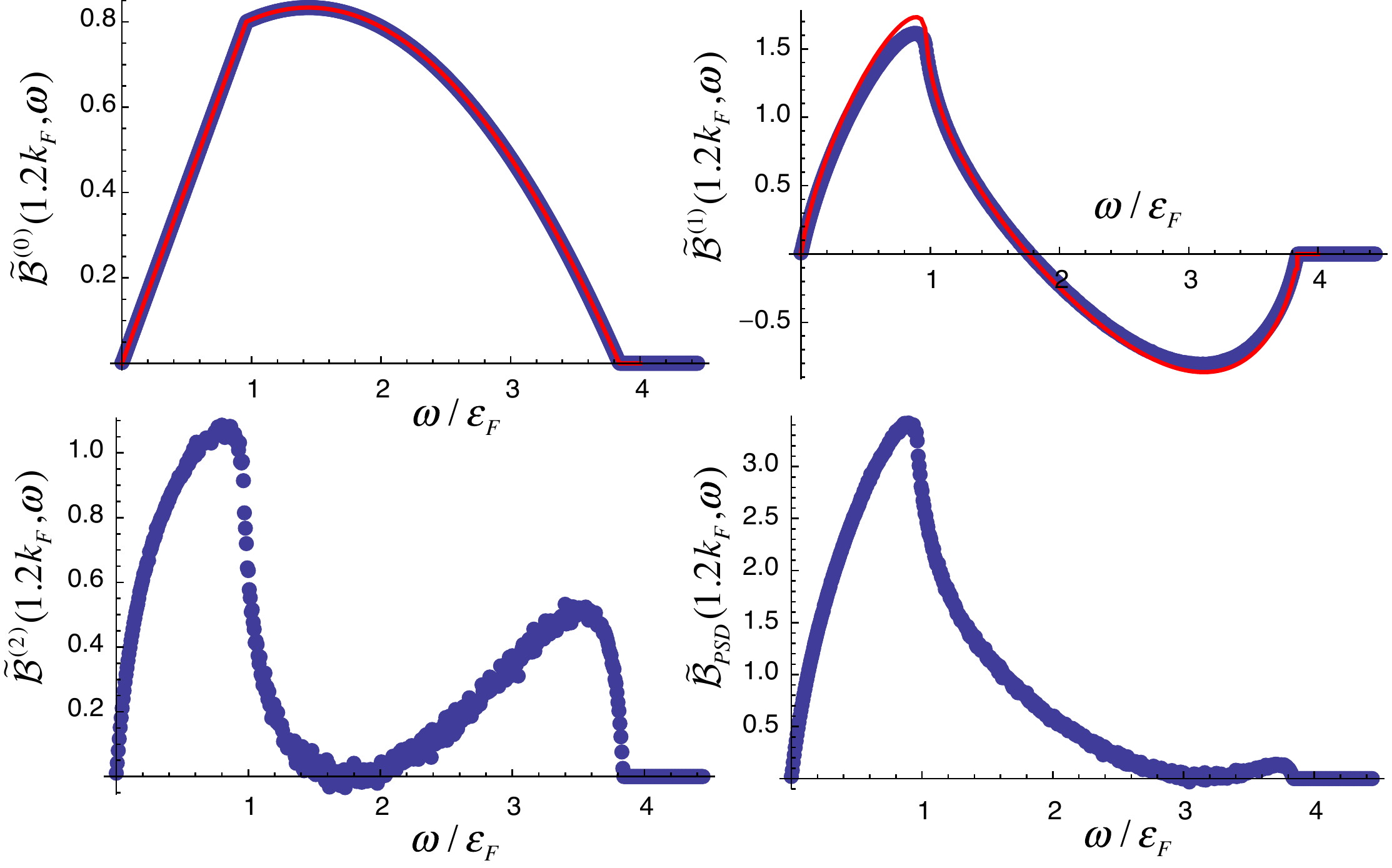}
\caption{(Color online) Cross-sections of the results in Fig.~\ref{fig:chi} for the
momentum value of $k=1.2k_F$ and density $r_s=3$. Top left: zeroth 
order contribution $\tilde{\mB}^{(0)}$. 
Top right: first order contribution $\tilde{\mB}^{(1)}$. Bottom left: second order
contribution $\tilde{\mB}^{(2)}$. Bottom right: the sum of all three contributions $\tilde{\mB}_{\rm PSD}$ 
which is positive.
Dots (blue) denote numerical Monte Carlo results. Solid lines (red) stand for analytical results.
 \label{fig:chi_k}}
\end{figure}

Let us look now at the results of the Monte-Carlo integration for a density of $r_s=3$.  The
full momentum ($0.1k_F\le k\le 2.5k_F$) and energy ($0\le \omega\le 12\epsilon_F$)
resolved spectral functions are shown in Fig.~\ref{fig:chi}. On this graph the numerically
produced points are projected on the analytically known results depicted as surfaces. For
the second order analytical expressions are not known, therefore, only points are
shown. At small momenta the spectral functions diverge, therefore we introduced some
truncation. The sum of all tree terms, $\tilde{\mB}_{\rm PSD}$, is always positive as an example of a cross-section
at $k=1.2k_F$ in Fig.~\ref{fig:chi_k} demonstrates. For some frequencies, however,
$\tilde{\mB}_{\rm PSD}(k,\omega)$ is rather small and lies within the error bar of the Monte Carlo
simulation.

%---------------------------%
%---------------------------%
%---------------------------%
%---------------------------%

The domain where $\tilde{\mB}^{(1)}(k,\omega)$ is negative is not 
bounded, see again
Fig.~\ref{fig:chi_dens_plot}. Thus, corrections originating from
$\tilde{\mB}^{(2)}(k,\omega)$ qualitatively modify the behavior of the spectral function
at any momentum. This, in view of the Hilbert transform in Eq.~\eqref{PR_rep}, leads to a
modification of the real part of the response function.  The general conclusion is that
the cutting procedure for PSD spectra works as it should.  The addition of the second
order vertex diagram correctly removes the negative parts of the response function to
first order in the vertex.

\subsection*{Cancellation between vertex corrections and self-energy insertions}

\begin{figure}[b!]
\centering \includegraphics[width=\columnwidth]{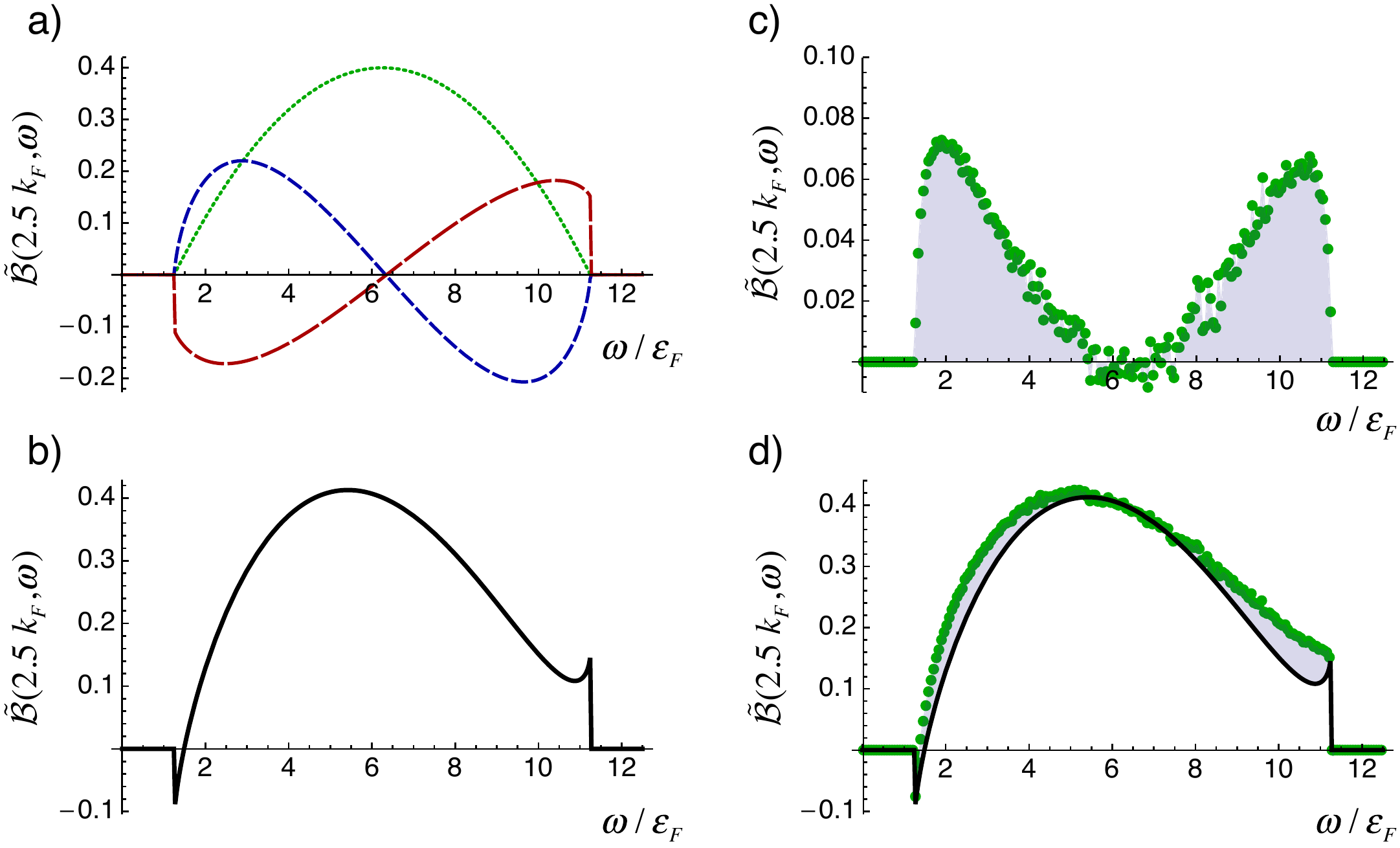}
\caption{(Color online) Spectral functions at momentum $k=2.5k_F$ and density $r_s=3$. (a)
  In addition to previously considered zeroth order ($\tilde{\mB}^{(0)}$, dotted) and
  first order ($\tilde{\mB}^{(1)}$, short dash) a contribution of the diagrams with
  self-energy insertion ($\tilde{\mB}^{(1,\mathrm{Se})}$, long dash) is shown. (b) Sum of
  three terms in panel (a).  (c) Second order contribution, $\tilde{\mB}^{(2)}$. (d)
  $\tilde{\mB}_{\rm PSD}$ including the first order self-energy diagrams (dots), solid
  line as in (b), shaded area denotes second order contribution. In order to cancel small
  negative spectral function at the edge of particle-hole continuum
  ($\omega=1.25\epsilon_F$) inclusion of more diagrams as shown at
  Fig.~\ref{fig:new_square} is required.
   \label{fig:SEdiagram}}
\end{figure}

\begin{figure}[t!]
\centering \includegraphics[width=\columnwidth]{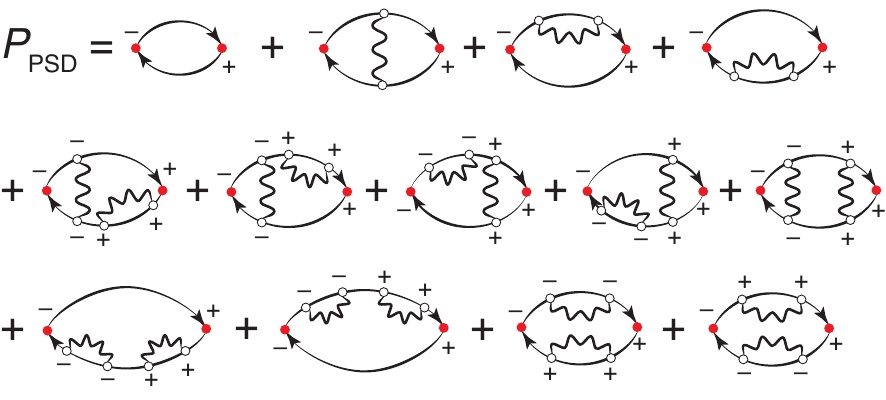}
\caption{(Color online) 
   $\mP_{\rm PSD}$ obtained from the cutting procedure including the first order self-energy diagrams. 
     \label{fig:new_square}}
\end{figure}

Although our example is the simplest one that illustrates the PSD  
diagrammatic theory, it is too
simple from a physical point of view. The main reason is that we used bare propagators
$G_0$ and bare interactions $v$. The first order vertex diagram is not the only first
order diagram as we missed two first order diagrams that contain 
exchange self-energy
insertions. If we had used an expansion in dressed Green's functions $G$ and dressed
interactions $W$ such diagrams would not appear as in that case we could restrict
ourselves to skeleton diagrams. However, in an expansion in $G_0$ and $v$ they become
relevant. In particular the self-energy diagrams lead to a cancellation of the divergent
small $k$-behavior of the spectral function $\tilde{\mB}^{(1)}(k,\omega)$.  For the case
$\omega=0$ this as been explicitly demonstrated by Engel and
Vosko~\cite{engel_wave-vector_1990}. It would therefore be a natural step to include the
first order self-energy diagrams as it would, for instance, guarantee 
the existence of a gradient expansion for the exchange-correlation 
energy, see below. Taken together these diagrams can be expanded in a
power series in terms of the momenta which plays an important role in determining the
gradient expansion of the exchange-correlation energy functional in density functional
theory.  It is well known that in the lowest order the gradient correction to the
exchange-correlation functional can be expressed
as~\cite{engel_wave-vector_1990,van_leeuwen_density_2013}
\[
\Delta E_{xc}[n_0,\delta n(\vec q)]=\frac12\int\frac{d^3q}{(2\pi)^3}\delta n(\vec q)
\frac{\Delta\mP(\vec q,0)}{[\mP(\vec q,0)]^2}\delta n(-\vec q),
\]
where $\delta n$ is the density variation with respect to the density $n_0$ of the
homogeneous system, and $\Delta\mP(\vec q,0)$ denotes corrections to the response function
from the first and higher order diagrams.  Therefore, the inclusion of $\tilde{\mB}^{(2)}$
will certainly modify the gradient expansion coefficients, e.g, due to Engel and
Vosko~\cite{engel_wave-vector_1990}. However, to have a well-defined gradient expansion
one has to add the self-energy diagrams as well.  Unfortunately, a simple addition would
destroy the positivity of the resulting spectral function again. This was noticed e.g. by
Brosens and Devreese~\cite{brosens_dynamical_1984} and is illustrated in
Fig.\ref{fig:SEdiagram}, where we also display the contribution of the self-energy diagram
($\tilde{\mB}^{(1,\mathrm{Se})}(k,\omega)$) calculated using the analytic expression of
Holas {\em et al.}\cite{holas_dynamic_1979}.
%---------------------------%
%---------------------------%
Therefore, if we desire to include the self-energy diagrams and still wish to guarantee
positivity we have to apply our PSD theory and consider an extended set of diagrams. The
minimal set that achieves this goal is displayed in Fig.~\ref{fig:new_square} which apart
from additional self-energy diagrams also contains mixed self-energy and vertex
diagrams. Rather than developing codes to evaluate these additional diagrams we found it
more worthwhile to explore approximations that involve dressed Green's functions and
interactions. First of all, the dressing of the interaction reduces the singular behavior
of the diagrams and secondly they reduce the number of diagrams to be evaluated since we
can then stick to skeletonic diagrams. However, since this requires an extensive
discussion by itself we will address this topic in a future publication. An alternative
route was undertaken by Brosens, Devreese and Lemmens in a series of
works~\cite{brosens_frequency-dependent_1976,brosens_frequency-dependent_1977,devreese_dielectric_1980,brosens_dielectric_1980}
using the variational solution of the linearized equation of motion for the electron
density distribution function. They have shown that the first term in a series expansion
of the variational result for the local field factor yields the lowest order diagrammatic
result. Since such solution also contains higher order terms the spectral function is
positive. However, it is not free from singularities suggesting again possible benefits of
working with dressed Green's functions and interactions.
 
%---------------------------%
%---------------------------%
%**************************************************************************************
\section{Conclusions and outlook}
\label{conclusions}
%**************************************************************************************

Vertex corrections in diagrammatic approximations to the polarizability are known to be
crucial for capturing double and higher particle-hole excitations, excitons, multiple
plasmon excitations, etc. as well as for estimating excitation life-times.  However, the
straighforward inclusion of MBPT vertex diagrams can lead to negative spectra, a drawback
which discouraged the scientific community to develop numerical recipes and tools for the
evaluation of these diagrams in molecules and solids. In this work we provided a simple
set of rules to select special combinations of diagrams yielding a positive spectrum.

In our formulation every MBPT diagram is written as the sum of partitions, and every
partition is cut into half-diagrams. We recognized that they are the half-diagrams the
fundamental quantities for a PSD expansion. In fact, the sum of squares of half-diagrams
corresponds to a special selection of partitions which is PSD by construction. The
requirement of positivity on the spectrum is important not only for the physical
interpretation of the results but also for the correct analytic structure of the
polarizability.  We demonstrated that a PSD polarizability cannot generate a (retarded)
density response function with poles in the upper-half of the complex frequency plane.
This is a critical property to converge self-consistent numerical schemes. Although the
PSD diagrammatic expansion put foward in this work applies equally well to bare as well as
dressed Green's functions a word of caution is due in the dressed case. The gluing of
skeletonic, i.e., self-energy insertion-free, half-diagrams can lead to nonskeletonic
polarizability diagrams. In order to avoid the double counting of some of the diagrams it
is therefore necessary to use Green's functions dressed with self-energy diagrams distinct
from those appearing in $\mP$.

A natural way to sum to infinite-order a subclass of polarizability diagrams is through
the BSE, an integral equation with kernel given by the functional derivative of the
self-energy with respect to the Green's function. Due to the popularity of the BSE we also
addressed the issue whether the polarizability which solves the BSE is PSD for conserving
self-energies, and found a negative answer.  The counter 
example is provided
by the self-energy in the second-Born or $GW$ approximations.  
Noteworthy these self-energies yield a
positive spectrum for the Green's function~\cite{stefanucci_diagrammatic_2014}; therefore
neither a conserving nor a PSD self-energy does necessarily generate a PSD polarizability
through the BSE.

The simplest approximation with vertex corrections is the first-order ladder diagram. This
diagram has been calculated both in finite and bulk systems and it is known to be not
PSD. How to include vertex corrections without altering the positivity of the spectrum has
been a long-standing problem, which we have solved in this work. By adding a partition of
the second-order ladder diagram we obtained the simplest PSD approximation with vertex
corrections. We then evaluated this approximation in the 3D 
homogeneous electron gas and confirmed numerically the correctness of our PSD theory. 
We stress again that the PSD property alone does not 
necessarily guarantee physically meaningful spectra. In fact, the PSD 
spectrum with vertex corrections has un unphysical divergency at zero 
frequency and momentum.
The inclusion of bubble diagrams with first-order exchange 
self-energy insertions removes this
 divergency but destroys the PSD property. We
worked out the minimal set of diagrams to turn this extended approximation into a PSD
one. The evaluation of the resulting extra diagrams is within reach of our code but it
requires a considerable numerical effort and it goes beyond the scope of the
present work. 

%**************************************************************************************
\section{Acknowledgments}
%**************************************************************************************
AMU would like to thank the Alfred Kordelin Foundation for support. 
GS acknowledges funding by MIUR FIRB Grant No. RBFR12SW0J. 
YP acknowledges support by the DFG through SFB762.  
RvL would like to thank the Academy of Finland for support.

%*****************************************************************************************
\appendix 
\section{Numerical evaluations of the analytical expressions for the 
first order polarizability}
\label{ApB}
As in the rest of the text we measure the momentum and the energy in units of the Fermi
momentum $k_F=1/(\alpha r_s)$ and the Fermi energy $\epsilon_F=k_F^2/2$.  In older papers
$k_F^2$ as the energy unit was used~\cite{holas_dynamic_1979}.  This must be taken into
account when comparing. Also notice that Engel and Vosko measured momentum in terms of
$2k_F$. It is natural because the first order polarizability has a logarithmic singularity
at this point. The imaginary part of the dielectric function resulting from the first
order polarizability is given by
%=================
\be
\text{Im}\,\varepsilon(q,\omega)=\frac{8}{\pi}\frac{\alpha^2r_s^2}{k^4}
\left(F^\mathrm{Ex}\Big(k,\frac{\omega}{2}\Big)+F^\mathrm{Se}\Big(k,\frac{\omega}{2}\Big)\right),
\ee 
%=================
whereas the scaled spectral functions considered in Sec.~\ref{numsec} read
%=================
\bea
\tilde{\mB}^{(1)}(k,\omega)&=&\frac{4}{\pi}\frac{\alpha r_s}{k^2}F^\mathrm{Ex}\Big(k,\frac{\omega}{2}\Big),\\
\tilde{\mB}^{(1,\mathrm{Se})}(k,\omega)&=&\frac{4}{\pi}\frac{\alpha r_s}{k^2}F^\mathrm{Se}\Big(k,\frac{\omega}{2}\Big),
\eea 
%=================
with the function $F$ non-zero at two domains in $k-\omega$ plane (restricted by the
Heaviside $\theta$-functions):
%=================
\begin{multline}
F^\mathrm{Ex,Se}(k,\bar{\omega})=
\theta\Big[1-\Big(\frac{\bar{\omega}}{k}-\frac{k}{2}\Big)^2\Big]
\Phi^\mathrm{Ex,Se}\Big(\frac{\bar{\omega}}{k}-\frac{k}{2},k\Big)\\
-\theta\Big[1-\Big(\frac{\bar{\omega}}{k}+\frac{k}{2}\Big)^2\Big]
\Phi^\mathrm{Ex,Se}\Big(-\frac{\bar{\omega}}{k}-\frac{k}{2},k\Big),
\end{multline}
%=================
and in turn
%=================
\bea
\Phi^\mathrm{Se}(\nu,k)&=&kf_L\big(\big[(k+\nu )^2+\left(1-\nu ^2\right)\big]^{1/2}\big)\nn\\
&&\quad\quad\quad-(k+\nu ) f_L(k+\nu )+\nu f_L(\nu ),\quad\\
\Phi^\mathrm{Ex}(\nu,k)&=&-G_1(\nu)+G_2(\nu+k,1-\nu^2),
\eea
%=================
where $f_L$ is the Lindhard function
\[
f_L(z)=\frac{1}{2}+\frac{1-z^2}{4z}\log \left|\frac{z+1}{1-z}\right|,
\]
and $G_1$ is (cf. Eq.~(2.15) of Holas~\emph{et al.}in Ref.~[\onlinecite{holas_dynamic_1979}])
%=================
\bea
G_1(\nu)&=&\frac14(1-\nu^2)g\Big(\frac{1-\nu}{1+\nu}\Big)-\frac12\nu\\
&\times&\big((1-\nu)\log(1-\nu)+(1+\nu)\log(1+\nu)-2\log2\big),\nn
\eea
%=================
with $g(z)=\mathrm{Li}_2(-z)-\mathrm{Li}_2\big(-1/z\big)$ represented in terms of the
polylogarithm functions. The second function is more involved, it is given by the Hilbert
transform which has to be computed numerically:
%=================
\be
G_2(x,y)=-\frac14\int_{-1}^{1}d\xi\frac{T(\xi,x,y)}{\xi-x}.
\ee
%=================
If $|x|<1$ the simplest way to avoid singularity is to exclude a small
($|x_a-x_b|<\epsilon$) interval $x\in(x_a,x_b)\subset (-1,1)$ from the integration.
Finally, the $T(\xi,x,y)$ function is defined as (cf. Eqs.~(2.18-2.23) of Holas~\emph{et
  al.} in Ref.~[\onlinecite{holas_dynamic_1979}]):
%=================
\bea
T(\xi,x,y)&=&\left(\frac{a_1}{2 t}-\frac{a_1}{2}+\left(1-\xi ^2\right) (1-t)\right) \nn\\
&\times&\log \left( 2 t \left(1-\xi ^2\right)+a_1\right)
-y \log \left(a_1\right)\nn\\
&+&\left(1-\xi ^2\right) (t-\log (t)-1)
\eea
%=================
in terms of auxiliary functions ($\gamma=x^2+y$, $a_1=2(\xi-x)^2$,
$a_2=4(1-\xi^2)(\xi-x)^2$, $a_3=2\xi(\xi-x)-1$, $\lambda=\frac{a_2}{(\gamma+a_3)^2}$) and
%=================
\bea
t(\xi,x,y)=\left\{\begin{array}{lc} \frac{(\xi -x)^2}{a_3+\gamma }
&|\lambda|\le\epsilon,\\
\frac{a_3+\gamma}{\left(1-\xi ^2\right)}\frac{\sqrt{1+\lambda}-1}{2}&
|\lambda|>\epsilon.
\end{array}\right.
\eea
%=================
From the imaginary part of the polarization function the real part can be computed through
the Hilbert transform. For the static case $\omega=0$ we have:
%=================
\bea
\mathrm{Re}\chi^{(1)}(k,0)\!\!&=&\!\!\frac{2}{\pi^3k^2}\!\int_\epsilon^{k(k+2)}\frac{d\omega}{\omega}
\left[F^\mathrm{Ex}\big(k,\frac{\omega}{2}\big)+F^\mathrm{Se}\big(k,\frac{\omega}{2}\big)\right]\nn\\
\!\!&=&\!\!-\frac1{\pi^3}\left[a\left(\frac{k}{2}\right)+b\left(\frac{k}{2}\right)\right],
\label{Rchi1}
\eea
%=================
where for $a(q)$ and $b(q)$ there are analytic expressions due to Engel and
Vosko\cite{engel_wave-vector_1990}:
%=================
\bea
b(q)&=&\frac{1-q^2}{8 q^2}\log^2\left|\frac{1+q}{1-q}\right|
+\frac{q+1}{2q}\log|1+q|\nn\\
&-&\frac{1-q}{2q}\log|1-q|-\log |q|,\\
a(q)&=&\frac{\left(1-q^2\right)}{24 q^2} 
\int_0^q \frac{1-x^2}{x^2}\log^3\left|\frac{1+x}{1-x}\right| \, dx\nn\\
&-&\left(\frac{1-q^2}{16 q^2}\log\left|\frac{1+q}{1-q}\right|
+\frac{1}{8 q}\right)\nn\\
&\times&\int_0^q \frac{1-x^2}{x^2}\log^2\left|\frac{1+x}{1-x}\right|\, dx\nn\\
&-&\frac{\left(1-q^4\right)}{48 q^3} \log^3\left|\frac{1+q}{1-q}\right|-b(q).
\label{Rchi2}
\eea Numerically Eq.~\eqref{Rchi2} is much faster than the Hilbert
transform~\eqref{Rchi1}. However, it is good to know that both ways yield identical
results that also agree with our Monte-Carlo simulations.

%*****************************************************************************************
\end{document}